\documentclass{elsart}

\usepackage{graphicx}
\usepackage{psfrag}

\usepackage{amssymb}
\usepackage{amsmath}
\usepackage{dsfont}
\usepackage{rotating}

\newcommand{\CK}{Cauchy-Kovalewski }

\newcommand{\PNM}{P_NP_M}

\newcommand{\tens}[1]{{\mathbf{#1}}}

\newcommand{\CFL}{\textnormal{CFL}}

\newcommand{\halb}{\frac{1}{2}}

\newcommand{\WL}{\mathcal{W}_h^-}
\newcommand{\WR}{\mathcal{W}_h^+}

\newcommand{\be}{\begin{equation}}
\newcommand{\ee}{\end{equation}}
\newcommand{\bdm}{\begin{displaymath}}
\newcommand{\edm}{\end{displaymath}}
\newcommand{\bea}{\begin{eqnarray}}
\newcommand{\eea}{\end{eqnarray}}

\begin{document}

\begin{frontmatter}



\title{Very High Order $\PNM$ Schemes on Unstructured
Meshes for the Resistive Relativistic MHD Equations}


\author[UNITN]{Michael Dumbser}
\ead{michael.dumbser@ing.unitn.it}
\author[MIGP]{Olindo Zanotti}
\ead{zanotti@aei.mpg.de}
\address[UNITN]{Laboratory of Applied Mathematics, University of Trento \\ Via Mesiano 77, I-38100 Trento, Italy}
\address[MIGP]{Max-Planck-Institut f\"ur Gravitationsphysik, Albert Einstein Institut \\ Am M\"uhlenberg 1, D-14476 Golm, Germany}

\begin{abstract}
In this paper we propose the first better than second
order accurate method in space and time for the numerical
solution of the  resistive relativistic
magnetohydrodynamics (RRMHD) equations on unstructured
meshes in multiple space dimensions. The nonlinear
system under consideration is purely hyperbolic and
contains a source term, the one for the evolution of the
electric field, that becomes stiff for low values of the
resistivity. 

For the spatial discretization we propose to use high
order $\PNM$ schemes as introduced in \cite{Dumbser2008}
for hyperbolic conservation laws and a high order
accurate unsplit time discretization is achieved using
the element-local space-time discontinuous Galerkin
approach proposed in \cite{DumbserEnauxToro} 
for one-dimensional balance laws with
stiff source terms. The divergence free character of the
magnetic field is accounted for through the divergence
cleaning procedure of Dedner et al. \cite{Dedneretal}.

To validate our high order method we first solve some
numerical test cases for which exact analytical reference
solutions are known and we also show numerical
convergence studies in the stiff limit of the RRMHD
equations using $\PNM$ schemes from third to fifth order
of accuracy in space and time. We also present some
applications with shock waves such as a classical shock
tube problem with  different values for the conductivity
as well as a relativistic MHD rotor problem and the
relativistic equivalent of the Orszag-Tang vortex
problem.  
We have verified that the proposed method can handle
equally well the resistive regime and the stiff limit of
ideal relativistic MHD. For these reasons it provides a
powerful tool  for relativistic astrophysical simulations
involving the appearance of magnetic reconnection.

\end{abstract}
\end{frontmatter}


\section{Introduction}
\label{sec.introduction}

Although the assumption of infinite conductivity is 
often justified in astrophysics, there are nevertheless
situations in which neglecting the resistivity of the
plasma may lead to rather inaccurate or simply wrong 
conclusions. 
This is particularly the case for those physical systems
involving processes that present magnetic reconnection, 
such as 
in the magnetospheres of pulsars near the
Y-point, where the outermost magnetic field lines
intersect the equatorial plane \cite{Gruzinov2005},
\cite{Uzdensky2003}; 
or in soft gamma-ray 
repeaters where giant flares could be the explanation 
of the observed strongly magnetized and relativistic
ejection events \cite{Lyutikov2006}; 
or in extragalactic jets, where particle acceleration by
magnetic reconnection in electron-positron
plasmas is supposed to take place \cite{Jaroschek2004},
\cite{Schopper1998}, 
and also in active galactic nuclei, where Petschek
magnetic reconnection, associated with MHD turbulence,  
may generate violent releases of energy
\cite{diMatteo1998}.
Moreover, the presence of relativistic motion
makes resistive effects quantitatively and qualitatively
different from those encountered in the Newtonian regime.
For example, the relativistic reconnection of Petschek
type with non-strictly parallel reconnecting fields
produces a strong compression of the plasma and the
energy of the reconnecting field can be largely
propagated away in the form of a Poynting flux
\cite{Lyubarsky2005}. In addition, the reconnection rate
is also affected, and it is
roughly obtained by replacing the 
Alfv\'en wave velocity with
the speed of light in the corresponding formulas.

For all these reasons, and moreover because the question on whether
relativistic magnetic reconnection is an efficient
energy converter is still under debate (see the discussion in
\cite{Lyutikov2003} and \cite{Zenitani2009})
there is a strong interest in the numerical solution of 
the full system of RRMHD, by providing a single computational tool 
that can equally well handle situations of
low and high resistivity, as commonly encountered in all
realistic physical scenarios.
The  equations to solve are particularly challenging from the 
numerical point of view, since, as recently 
shown by \cite{Komissarov2007} and \cite{Palenzuela2009}, 
they become stiff for large values of the conductivity.
Namely, the RRMHD equations can be cast into the following general 
form of a hyperbolic balance law 
\begin{equation}
\label{eqn.pde.nc}
\frac{\partial}{\partial t} W + \nabla \cdot
\tens{F}\left(W\right) = S(W),  
\end{equation}
where $W$ is the state vector, $\tens{F}(W)$
is a nonlinear flux tensor that depends on the state $W$
and  $S(W)$ is a nonlinear source term that becomes stiff at 
high conductivities.
In this paper we solve the RRMHD equations by applying the high 
order accurate method recently proposed by \cite{DumbserEnauxToro} 
to cope with stiff source terms on the right hand side of (\ref{eqn.pde.nc})
and maintaining at the same time better than second order of accuracy
in space and time.  
The numerical method is formulated as one-step local predictor 
global corrector method. The predictor is based on an element-local 
weak solution of \eqref{eqn.pde.nc}, where inside each element the 
governing PDE \eqref{eqn.pde.nc} is solved \textit{in the small} 
(see \cite{eno})
by means of a locally implicit space-time discontinuous Galerkin scheme. 
This leads to an algebraic system of non-linear equations that must 
be solved individually for each element. The globally explicit update 
in time, on the other hand, is obtained by either standard finite volume 
or discontinuous Galerkin methods, or, finally, by a recently proposed
generalization of the two named $\PNM$ schemes according to 
\cite{Dumbser2008}.

The plan of the paper is as follows. 
In Sect.~\ref{sec.rrmhd}
we briefly review the peculiar features of the RRMHD equations. 
The core of the numerical method is described in
Sect.~\ref{sec.numethod} while Sect.~\ref{sec.appl} is
devoted to the validation of the scheme through  a large
class of numerical tests.  
Finally, the conclusions are reported
in Sect.~\ref{sec.conclusion}.
We have considered only flat spacetimes in Cartesian
coordinates, namely the metric 
$\eta_{\mu\nu}={\rm diag}(-1,1,1,1)$, where from now onwards 
we agree to use Greek letters
$\mu,\nu,\lambda,\ldots$ (running from 0 to 3) for
indices of four-dimensional space-time tensors, while
using Latin letters $i,j,k,\ldots$ (running from 1 to 3) 
for indices of three-dimensional spatial tensors.
Finally, we set the speed of light $c=1$ and make use of the
Lorentz-Heaviside notation for the electromagnetic
quantities, such that all $\sqrt{4\pi}$ factors disappear.  
We use Einstein summation convention over repeated indices.

\section{The Resistive Relativistic MHD Equations}
\label{sec.rrmhd}

\subsection{Conservative Formulation as a Stiff Hyperbolic Balance Law}
\label{sec.rrmhd.formulation}
%
The total energy-momentum tensor of the system is made up by two
contributions,
$T^{\mu\nu}=T^{\mu\nu}_{m}+T^{\mu\nu}_{f}$. The first one
is due to matter 
\be
T^{\mu\nu}_{m}=\omega\,u^{\,\mu}u^{\nu} + p\eta^{\,\mu\nu},
\label{eq:T_matter}
\ee
where $u^\mu$ is the four velocity of the fluid, while
$\omega$ and $p$ are  the enthalpy and the pressure as
measured in the co-moving frame of the fluid. The second
contribution is due to the electromagnetic field
\be
T^{\mu\nu}_{f}={F^{\mu}}_{\lambda}F^{\nu\lambda}-\textstyle{\frac{1}{4}}(F^{\lambda\kappa}F_{\lambda\kappa})\eta^{\,\mu\nu}
\ ,
\label{eq:T_field}
\ee
where $F^{\mu\nu}$ is the electromagnetic tensor. If we
introduce a laboratory observer defined by a four
velocity $n^\mu=(1,0,0,0)$, then the fluid four velocity
$u^\mu$ and the standard three velocity in the laboratory
frame are related as $\vec v = v^i=u^i/\Gamma$, where
$\Gamma=({1-\vec v^{\, 2}})^{-0.5}$ is the 
Lorentz factor of the fluid with respect to the
laboratory frame.
The  electromagnetic tensor, on the other hand, can be written as 
\be
F^{\mu\nu} = n^{\,\mu}E^{\nu} - E^{\mu}n^{\nu} +
\epsilon^{\,\mu\nu\lambda\kappa}B_{\lambda}n_{\kappa}, 
\ee
where $E^{\nu}$ and  $B^{\nu}$ are the two spatial vectors
($E^0=B^0=0$, $E^i=E_i$, $B^i=B_i$) representing electric and
magnetic field, respectively, while
$\epsilon^{\,\mu\nu\lambda\kappa}=[\mu\nu\lambda\kappa]$
is the completely antisymmetric spacetime Levi-Civita
tensor, with the convention that $\epsilon^{\, 0 1 2 3}=1$.
%
The equations of motion can be derived from the
conservation laws
\be
\partial_\mu T^{\mu\nu}=0
\ee
and from the continuity equation
\be
\partial_\mu(\rho u^\mu)=0,
\ee
where $\rho$ is the rest mass density of the fluid. The
electromagnetic field, on the other hand, obeys the
Maxwell equations expressed in the form
\be
\partial_{\mu}F^{*\mu\nu} = 0, 
\qquad 
\partial_{\nu}F^{\mu\nu} = I^{\mu}, 
\label{eq:maxwell}
\ee
%
where
$F^{*\mu\nu}=\frac{1}{2}\epsilon^{\,\mu\nu\lambda\kappa}F_{\lambda\kappa}$
is the dual of the electromagnetic tensor, while $I^\mu$
is the four vector of electric currents. 
The equations of resistive MHD differ from those of ideal
MHD mainly because the second couple of Maxwell equations
(\ref{eq:maxwell}),  accounting for the time evolution
and for the divergence of the  electric field, need  to be explicitly
solved. This  requires that a relation is given between
the currents and the electromagnetic field, the so called
Ohm's law. In its most general form, the relativistic
formulation of Ohm's law is a non linear propagation equation
\cite{Kandus2008}, but here, as in \cite{Komissarov2007}, we
will simply assume that
\be
\label{ohm1}
I^\mu=q_0 u^\mu + \sigma F^{\mu\nu}u_\nu, 
\ee
where $\rho_c$ is the charge density in the co-moving
frame while $\sigma$ is the electric conductivity. 
From (\ref{ohm1}) we easily derive the following
expression for the spatial current vector
\be
\vec J = \rho_c \vec v + \sigma \Gamma [\vec E + \vec v \times
\vec B - (\vec E\cdot \vec v)\vec v] \ ,
\ee
where $\rho_c$ is the charge density in the laboratory
frame. As done by \cite{Komissarov2007} and
\cite{Palenzuela2009}, to whom we address the reader for
further details, we take care of the divergence-free 
character 
of the magnetic field by adopting the {\em divergence cleaning
approach} presented in \cite{Dedneretal}, namely by
introducing two additional scalar fields $\Psi$ and
$\Phi$ that propagate away the deviations of the
divergences of the electric and of the magnetic fields
from the values prescribed by Maxwell's equations.
In total, the full set of RRMHD equations
include the five equations for the
fluid, plus the six equations for the evolution of the
electric and of the
magnetic field, plus the two equations about the
divergences of the two fields, plus one more equation
expressing the conservation of the total charge. 
In Cartesian coordinates, using the abbreviations 
$\partial_t = \frac{\partial}{\partial t}$ and 
$\partial_i = \frac{\partial}{\partial x_i}$, they can 
be written as: 
\vspace{-10mm}
\bea
\label{fluid1}
&& \partial_t D + \partial_i (D v^i)=0, \\
\label{fluid2-4}
&&\partial_t S_j + \partial_i Z_{j}^i=0, \\
\label{fluid5}
&&\partial_t \tau + \partial_i S^i=0, \\
\label{electric6-8}
&&\partial_t E^i - \epsilon^{ijk}\partial_j B_k + \partial_i
\Psi = -J^i, \\
&&\partial_t B^i + \epsilon^{ijk}\partial_j E_k + \partial_i
\Phi = 0, \\
\label{divE}
&&\partial_t \Psi + \partial_i E^i = \rho_c - \kappa \Psi,
\\
\label{divB}
&&\partial_t \Phi + \partial_i B^i = - \kappa \Phi, \\
\label{charge}
&&\partial_t \rho_c + \partial_i J^i = 0, 
\eea
where the conservative variables of the fluid are
\vspace{-10mm} 
\bea
D&=&\rho \Gamma, \\
S^i & = & \omega \Gamma^2 v^i + \epsilon^{ijk}E_jB_k, 
\label{eq:S} \\
\tau & = & \omega \Gamma^2 - p +
\textstyle{\frac{1}{2}}(E^2+B^2) \ ,
\label{eq:U}
\eea
expressing, respectively, the relativistic mass density, the
momentum density and the total energy density.
The spatial tensor $Z^i_j$ in (\ref{fluid2-4}), representing the momentum 
flux density, is 
\vspace{-10mm} 
\bea
\label{eq:W} 
Z^i_j & = & \omega \Gamma^2 v^i\,v_j -E^i\,E_j-B^i\,B_j+\left[p+\textstyle{\frac{1}{2}}(E^2+B^2)\right]\,\delta^i_j,
\eea
where $\delta^i_j$ is the Kronecker delta.
In the rest of the paper we have assumed the equation of
state of an ideal gas, namely
\be
\label{eos}
p=(\gamma-1) \rho\epsilon=\gamma_1(\omega-\rho), 
\ee
where $\gamma$ is the adiabatic index,
$\gamma_1=(\gamma-1)/\gamma$, $\epsilon$ is the specific internal energy and
$\omega = \rho \epsilon + \rho + p$ is the enthalpy.  
The system of equations (\ref{fluid1})-(\ref{charge}) is
written as a hyperbolic system of balance laws as in (\ref{eqn.pde.nc}), it has
source terms in the three equations (\ref{electric6-8})
that are potentially stiff (see \cite{Palenzuela2009} for a
more detailed description of the different limits of the
resistive MHD equations) and, as such, it can be treated
with the procedure proposed by \cite{DumbserEnauxToro},
as we will show in Sect.~\ref{sec.numethod}. 

\subsection{Closed form recovering of the primitive variables
from the conservative ones
}
\label{sec.rrmhd.consprim}
A fundamental difference with respect to the ideal MHD
case is that the augmented set of conservative variables of
the resistive relativistic equations allows for the
recovering of the primitive variables from the
conservative ones in closed form, 
at least when the equation of state is that of an ideal gas.
This can be seen in the following way. Firstly,  
we shift the cross pruduct $\vec{E}\times\vec{B}$
from the right hand side to the left hand side 
of Eq.~(\ref{eq:S}), then we square it, and we obtain
\be
\label{eq:S_bis}
(\vec{S}-\vec{E}\times\vec{B})^2 =\omega^2 \Gamma^2 (\Gamma^2-1) \ .
\ee
On the other hand, from (\ref{eq:U})
we obtain the enthalpy $\omega$ as 
\be
\label{enthalpy}
\omega=\frac{\tau - \frac{1}{2}(E^2+B^2) -\gamma_1 D/\Gamma}{\Gamma^2-\gamma_1} \ ,
\ee
where we have used $p=\gamma_1(\omega - \rho)$ as in (\ref{eos}).
After replacing
(\ref{enthalpy}) into (\ref{eq:S_bis}), simple
calculations lead to 
the following quartic equation in the unknown
Lorentz factor $\Gamma$ as follows: 
\be
\label{quartic}
A_4 \Gamma^4 + A_3 \Gamma^3 + A_2 \Gamma^2 + A_1 \Gamma +
A_0 = 0,  
\ee
where
\be
A_4 = C_1-C_2^2, \qquad 
A_3 = 2 C_2\gamma_1 D, \qquad   
A_2 = C_2^2-2 C_1\gamma_1 - \gamma_1^2 D^2, 
\ee 
\be 
A_1 = -2C_2\gamma_1 D, \qquad 
A_0 = \gamma_1^2(C_1 + D^2),
\ee 

with $C_1\equiv(\vec{S}-\vec{E}\times\vec{B})^2$, and 
$C_2\equiv \tau - \frac{1}{2}(E^2+B^2)$. The quartic
(\ref{quartic}) can be solved either analytically using the
approach of Ferrari and Cardano \cite{ArsMagna} or numerically via a 
Newton-Raphson scheme. In our numerical experiments we found that for the 
purpose of accuracy and robustness, it is advisable to solve the quartic 
first analytically and then to improve the accuracy of the result by 
one or two additional Newton iterations. This is necessary since the computations of the 
roots for the analytical solution of the quartic may introduce a significant amount of 
roundoff errors on finite precision computer hardware even when using double precision 
arithmetic. It is only for this reason that the additional Newton iterations are 
performed. This step would not be necessary with exact arithmetic. \\
As already pointed out by \cite{Zenitani2009}, it turns out that (\ref{quartic}) 
has two complex conjugate solutions, plus two real solutions, of which only one 
is larger than unity, as physically required.   
Once the Lorentz factor is known, the other primitive
variables can be computed in a straightforward manner. 

\section{Numerical Method}
\label{sec.numethod} 

\subsection{The $\PNM$ Reconstruction Operator on Unstructured Meshes} 
\label{sec.reconstruction}

The main ingredient of the proposed numerical method to reach high order of accuracy in space is the 
$\PNM$ reconstruction operator on unstructured meshes first introduced in \cite{Dumbser2008}. It is a 
direct extension of the reconstruction algorithm proposed in \cite{DumbserKaeser06b,DumbserKaeser07} 
for finite volume schemes. For the details, we refer to the above mentioned publications and give only
a short review in this section. The computational domain $\Omega$ is discretized by conforming elements 
$Q_i$, indexed by a single mono-index $i$ ranging from 1 to the total number of elements $N_E$. The 
elements are chosen to be triangles in 2D and tetrahedrons in 3D. The union of all elements is called 
the triangulation or tetrahedrization of the domain, respectively, 
\begin{equation}
\label{eqn.tetdef}
 \mathcal{Q}_{\Omega} = \bigcup \limits_{i=1}^{N_E} Q_i.
\end{equation}
At the beginning of a time-step, the numerical solution of \eqref{eqn.pde.nc} for the state vector $W$,
denoted by $u_h$, is represented by piecewise polynomials of degree $N$ from the space $V_h$, spanned 
by the basis functions $\Phi_l=\Phi_l(\vec x)$, i.e. at $t=t^n$ we have for each element 
\begin{equation}
\label{eqn.ansatz.u} 
  u_h(\vec x,t^n) = \sum_l \Phi_l(\vec x) \hat u_l^n.
\end{equation} 
From the polynomials $u_h$, we then reconstruct piecewise polynomials $w_h$ of degree 
$M \geq N$ from the space $W_h$, spanned by the basis functions $\Psi_l=\Psi_l(\vec x)$:
\begin{equation}
\label{eqn.ansatz.w} 
  w_h(\vec x,t^n) = \sum_l \Psi_l(\vec x) \hat w_l^n.
\end{equation}  
As stated in \cite{Dumbser2008}, the $\Psi_l$ form an orthogonal basis and are identical with the $\Phi_l$ up to polynomial 
degree $N$. For performing the reconstruction on element $Q_i$, we now choose a reconstruction stencil 
\begin{equation}
\label{eqn.stencildef} 
 \mathcal{S}_i = \bigcup \limits_{k=1}^{n_e} Q_{j(k)} 
\end{equation}
that contains a total number of $n_e$ elements. Here $1\leq k \leq n_e$ is a local index, counting the 
elements in the stencil, and $j=j(k)$ is the mapping from the local index $k$ to the global indexation 
of the elements in $\mathcal{Q}_{\Omega}$. 
For ease of notation, we write in the following only $j$, meaning $j=j(k)$. 

In the present paper the three operators 
\begin{equation}
 \label{eqn.operators1}
  \left<f,g\right>_{Q_i} = 
      \int \limits_{t^n}^{t^{n+1}} \int \limits_{Q_i} \left( f(\vec x, t) \cdot g(\vec x, t) \right) d V \, d t, 
\end{equation}
\begin{equation}
  \left[f,g\right]^{t}_{Q_i} = 
      \int \limits_{Q_i} \left( f(\vec x, t) \cdot g(\vec x, t) \right) d V, 
\end{equation} 
\begin{equation}
  \left\{f,g\right\}_{\partial Q_i} =  
      \int \limits_{t^n}^{t^{n+1}} \int \limits_{\partial Q_i} \left( f(\vec x, t) \cdot g(\vec x, t) \right) d S \, d t, 
\end{equation}      
denote the scalar products of two functions $f$ and $g$ over the space-time element 
$Q_i \times \left[t^n;t^{n+1}\right]$, over the spatial
element $Q_i$, and over the space-time boundary element 
$\partial Q_i \times \left[t^n;t^{n+1}\right]$ respectively. The operators $\left<f,g\right>$ and $\left[f,g\right]^t$, 
written without the index $Q_i$, will denote scalar products on the space-time reference element $Q_E \times [0;1]$ and on the spatial reference element $Q_E$ at time $t$, respectively. The spatial reference element $Q_E$ is defined as the unit 
simplex with vertices $(0,0)$, $(1,0)$, $(0,1)$ in two space dimensions and vertices $(0,0,0)$, $(1,0,0)$, $(0,1,0)$ and $(0,0,1)$ in three space dimensions, respectively.    

The reconstruction is now obtained via $L_2$-projection of the (unknown) piecewise polynomials $w_h$ from 
the space $W_h$ into the space $V_h$ on each stencil $ \mathcal{S}_i$ as follows:
\begin{equation}
 \label{eqn.pnmrec}
  \left[ \Phi_k, w_h \right]^{t^n}_{Q_j} = \left[ \Phi_k, u_h \right]^{t^n}_{Q_j}, \qquad \forall Q_j \in \mathcal{S}_i. 
\end{equation}

Note that during the reconstruction step, the polynomials $w_h$ are continuously extended over the 
whole stencil $\mathcal{S}_i$. After reconstruction, the piecewise polynomials $w_h$ are again 
restricted onto each element $Q_i$. 
The number of elements in the stencils are chosen in such a way that the number of equations in
\eqref{eqn.pnmrec} is \textit{larger} that the number of degrees of freedom in the space $W_h$. 
Eqn. \eqref{eqn.pnmrec} constitutes thus an overdetermined linear algebraic equation system for the coefficients of $w_h$ and 
is solved using a constrained least squares technique, see \cite{Dumbser2008,DumbserKaeser06b}. 
The linear constraint is that Eqn. \eqref{eqn.pnmrec} is at least exactly satisfied for $Q_j=Q_i$, i.e. 
inside the element $Q_i$ under consideration. The integral on the left hand side in \eqref{eqn.pnmrec} 
is computed using classical multidimensional Gaussian quadrature of appropriate order, see \cite{stroud}. 
The integral on the right hand side can be computed analytically and involves the standard element 
mass-matrix.  \\
The resulting $M$-exact $\PNM$ least squares reconstruction can be interpreted as a generalization of the 
$k$-exact reconstruction proposed for pure finite volume schemes by Barth and Frederickson in their 
pioneering work \cite{barthlsq}. 

\subsection{The Local Space-Time Discontinuous Galerkin Predictor for Stiff Balance Laws} 
\label{sec.galerkin}

The original ENO scheme of Harten et al. \cite{eno} as well as the ADER-FV and ADER-DG schemes 
developed by Titarev and Toro \cite{titarevtoro} and Dumbser and Munz \cite{dumbser_jsc} use the
governing PDE itself in its strong differential form to obtain high order of accuracy in time. 
This is achieved via the so-called \CK procedure that substitutes time derivatives with space
derivatives via successive differentiation of the governing PDE with respect to space and time. 
This procedure becomes very cumbersome or even impossible for general nonlinear hyperbolic PDE
systems. In \cite{DumbserEnauxToro}, \cite{Dumbser2008} and \cite{BalsaraRumpf} an fully numerical 
approach was presented that replaces the semi-analytical \CK procedure by a 
\textit{local weak formulation} of the governing PDE in space-time. 
While the approach presented in \cite{DumbserEnauxToro} relies on a local discontinuous Galerkin 
approach in space-time, which is also able to handle stiff source terms, the methods proposed in 
\cite{Dumbser2008} uses a local continuous Galerkin method in space-time as predictor. In the 
present article we use the local space-time discontinuous Galerkin method due to the stiffness of the 
source terms. \\
We underline that the local space-time DG method is only used as \textit{local predictor} for the 
construction of a solution of the PDE \textit{in the small}, as it was called by Harten et al. in \cite{eno}. 
The local space-time predictors are then inserted into a \textit{global corrector}, which is fully explicit 
and provides the coupling between neighbor cells. As a consequence, the resulting nonlinear algebraic systems 
of the local space-time Galerkin methods are element local and not globally coupled, as in the global and
fully implicit space-time Galerkin approach introduced by van der Vegt and van der Ven 
\cite{spacetimedg1,KlaijVanDerVegt}. \\ 

The details of the local space-time DG predictor method are already given in \cite{DumbserEnauxToro}, 
hence we will only briefly recall the basic ideas here. We start from the strong formulation of PDE 
\eqref{eqn.pde.nc} and transform the PDE into the reference coordinate system $(\vec \xi, \tau)$ of the 
space-time reference element $Q_E \times [0;1]$ with $\vec \xi = (\xi, \eta)$ and $\nabla_\xi$ being the 
nabla operator in the $\xi-\eta$ reference system:    
\begin{equation}
\label{eqn.pde.nc.2d} 
 \frac{\partial}{\partial \tau} W 
    + \nabla_\xi \cdot \tens{F}^* \left( W \right) = S^*. 
\end{equation}
The modified flux tensor and the modified source term are given by 
\begin{equation}
  \tens{F}^* := \Delta t \, \tens{F}(W) J^T, \quad  
  S^* := \Delta t S(W),  \quad  
  J = \frac{\partial \vec \xi}{\partial \vec x},
\end{equation}
as revealed by simple algebraic manipulations. We now multiply Eqn. \eqref{eqn.pde.nc.2d} by a space-time test 
function $\theta_k=\theta_k(\xi,\eta,\tau)$ from the space of piecewise space-time polynomials of
degree $M$ and integrate over the space-time reference control volume $Q_E \times [0;1]$ to obtain 
the following weak formulation: 
\begin{equation}
\label{eqn.pde.nc.weak1} 
 \left< \theta_k, \frac{\partial}{\partial \tau} \mathcal{W}_h \right> 
    + \left< \theta_k, \nabla_\xi \cdot \tens{F}_h^* \left(\mathcal{W}_h\right) \right> 
    = \left< \theta_k, \mathcal{S}^*_h \left( \mathcal{W}_h \right) \right>. 
\end{equation}
In the numerical solution of Eqn. \eqref{eqn.pde.nc.weak1} 
we use the same ansatz for 
$\mathcal{W}_h$ as well as for the flux tensor and the source
term , i.e.
\begin{equation}
\label{eqn.st.state} 
 \mathcal{W}_h = \mathcal{W}_h(\xi,\eta,\tau) = 
 \sum \limits_l \theta_l(\xi,\eta,\tau) \widehat{\mathcal{W}}_l := \theta_l \widehat{\mathcal{W}}_l,
\end{equation}
\begin{equation}
\label{eqn.st.flux} 
 \mathcal{\tens{F}}^*_h = \mathcal{\tens{F}}^*_h(\xi,\eta,\tau) = 
 \sum \limits_l \theta_l(\xi,\eta,\tau) \widehat{\mathcal{\tens{F}}^*_l} := \theta_l \widehat{\mathcal{\tens{F}}^*_l},
\end{equation}
\begin{equation}
\label{eqn.st.source} 
 \mathcal{S}^*_h = \mathcal{S}^*_h(\xi,\eta,\tau) = 
 \sum \limits_l \theta_l(\xi,\eta,\tau)
 \widehat{\mathcal{S}^*_l} := \theta_l
 \widehat{\mathcal{S}^*_l} \ .
\end{equation}
The degrees of freedom of the flux $\widehat{\mathcal{\tens{F}}^*_l}$ and the source $\widehat{\mathcal{S}^*_l}$ can be 
computed from the ones of the state vector $\widehat{\mathcal{W}}_l$ either via the more accurate but also more expensive $L^2$-projection,  
\begin{equation}
  \left< \theta_l, \theta_l \right> \widehat{\mathcal{\tens{F}}^*_l} = \left< \theta_k, \mathcal{\tens{F}}^*(\mathcal{W}_h) \right>, 
 \qquad 
  \left< \theta_l, \theta_l \right> \widehat{\mathcal{{S}}^*_l} = \left< \theta_k, \mathcal{{S}}^*(\mathcal{W}_h) \right>, 
\end{equation}
or in a simple and cheap nodal fashion, if a nodal space-time basis as the one in \cite{Dumbser2008} is used:  
\begin{equation}
  \widehat{\mathcal{\tens{F}}^*_l} = \mathcal{\tens{F}}^*(\widehat{\mathcal{W}}_l), 
 \qquad 
  \widehat{\mathcal{{S}}^*_l} = \mathcal{{S}}^*(\widehat{\mathcal{W}}_l).  
\end{equation} 
In the element-local weak formulation of Eq.~\eqref{eqn.pde.nc.weak1} we apply integration by parts to the first term 
and thus obtain 
\begin{equation}
\label{eqn.pde.nc.dg1} 
   \left[ \theta_k, \mathcal{W}_h \right]^1 - \left[ \theta_k, w_h \right]^0 
   - \left< \frac{\partial}{\partial \tau} \theta_k, \mathcal{W}_h  \right>      
    + \left< \theta_k, \nabla_\xi \cdot \mathcal{\tens{F}}^*_h \right> 
    = \left< \theta_k, \mathcal{S}^*_h  \right>, 
\end{equation}  
where the initial condition at relative time $\tau=0$ is taken into account in a \textit{weak sense} by the term 
$\left[ \theta_k, w_h \right]^0$. We recall that $w_h$ is the piecewise polynomial obtained by the high order $\PNM$ 
reconstruction operator summarized in
\ref{sec.reconstruction}. We also note that the first two terms in \eqref{eqn.pde.nc.dg1} 
correspond to the choice of an upwind flux in the time direction, which is consistent with the causality principle that 
states that no effect can occur before its cause. Inserting the ansatz \eqref{eqn.st.state}-\eqref{eqn.st.source} 
into \eqref{eqn.pde.nc.dg1} we obtain
\begin{equation}
\label{eqn.pde.nc.dg2} 
   \left( \left[ \theta_k, \theta_l \right]^1 - \left< \frac{\partial}{\partial \tau} \theta_k, \theta_l \right> \right)
    \widehat{\mathcal{W}_l} 
   - \left[ \theta_k, \Psi_l \right]^0 \hat w_l^n 
    + \left< \theta_k, \nabla_\xi \theta_l \right> \cdot \widehat{\mathcal{\tens{F}}}^*_l 
    = \left< \theta_k, \theta_l \right>
    \widehat{\mathcal{S}}^*_l \ .
\end{equation}
After defining the following universal matrices (that need to be computed only once on the reference element) 
$\mathbf{K_1} = \left[ \theta_k, \theta_l \right]^1 - \left< \frac{\partial}{\partial \tau} \theta_k, \theta_l \right> $,
$\mathbf{K_{\xi}} = \left< \theta_k, \nabla_\xi \theta_l \right> $,
$\mathbf{M} = \left< \theta_k, \theta_l \right>  $,
$\mathbf{F_0} = \left[ \theta_k, \Psi_l \right]^0$
we can rewrite \eqref{eqn.pde.nc.dg2} in the more compact matrix notation:  
\begin{equation}
\label{eqn.pde.nc.dg3} 
   \mathbf{K_1} \widehat{\mathcal{W}_l} + \mathbf{K_{\xi}} \cdot \widehat{\mathcal{\tens{F}}}^{*}_l  
    = \mathbf{F_0} \hat w_l^n + \mathbf{M} \, \widehat{\mathcal{S}}^{*}_l.  
\end{equation}
Eqn. \eqref{eqn.pde.nc.dg3} is an element-local nonlinear algebraic system for the unknowns $\widehat{\mathcal{W}_l}$. 
For its solution we use the following simple iterative scheme, similar to the one proposed 
in \cite{Dumbser2008}:     
\begin{equation}
\label{eqn.pde.nc.dg4} 
    \widehat{\mathcal{W}_l}^{i+1} - (\mathbf{K_1})^{-1} \mathbf{M} \widehat{\mathcal{S}}^{*,i+1}_l = 
    (\mathbf{K_1})^{-1} \mathbf{F_0} \hat w_l^n - (\mathbf{K_1})^{-1} \mathbf{K_{\xi}} \cdot \widehat{\mathcal{\tens{F}}}^{*,i}_l. 
\end{equation}
As in \cite{Dumbser2008} the matrices contained in $(\mathbf{K_1})^{-1} \mathbf{K_{\xi}}$ have the remarkable property that  \textbf{all} their eigenvalues are \textbf{zero}, which makes \eqref{eqn.pde.nc.dg4} a contractive fixed point iteration in 
the homogeneous case (i.e. when $S=0$) and thus existence, uniqueness and convergence to the unique solution are guaranteed by the 
Banach fixed point theorem. Furthermore, in the linear case, the method is even guaranteed to converge to the exact solution in 
$M+1$ steps from any initial guess. In the non-homogeneous case with stiff source terms, however, it is necessary to take the
source implicitly, which is done in the present paper. We use the following simplified model for the implicit source term: 
\begin{equation}
  \widehat{\mathcal{S}}^{*,i+1}_l \approx \widehat{\mathcal{S}}^{*,i}_l + \Delta t \frac{\partial S}{\partial W} \left( \widehat{\mathcal{W}_l}^{i+1} - \widehat{\mathcal{W}_l}^{i} \right), \quad 
  \frac{\partial S}{\partial W} = \frac{\partial S}{\partial V} \frac{\partial V}{\partial W},  \quad 
  \frac{\partial V}{\partial W} = \left( \frac{\partial W}{\partial V} \right)^{-1}, 
\end{equation} 
where the Jacobian of the source with respect to the conservative variables $W$ is computed by the chain rule, taking first the derivatives with respect to the vector of primitive variables $V$. The derivative of $V$ with respect to $W$ can be computed easily
by the theorem on the derivative of the inverse function. To simplify the computations, we evaluate $\frac{\partial S}{\partial W}$ only once per iteration at the current space-time average value of $\mathcal{W}_h$. 

In our numerical experiments we also found that in the very stiff case, the choice of the initial guess  $\widehat{\mathcal{W}_l}^{0}$ seems to be very crucial. We therefore adopt the following strategy:
First we solve \eqref{eqn.pde.nc.dg4} at the first order level, which becomes a simple Newton-Raphson scheme for the space-time  cell-average $\overline{\mathcal{W}}$ as 
\begin{equation}
\label{eqn.pde.nc.dgO1} 
    \bar{f} = \overline{\mathcal{W}} - S^*\left( \overline{\mathcal{W}} \right) - \bar{u}_i^n = 0, 
\end{equation}
where the initial guess $\overline{\mathcal{W}}^{0} = \bar{u}_i^n$ is used for all variables apart from the electric
field. For the electric field we use $\vec E$ obtained from the relaxation of $\vec E$ and $\vec v$ to equilibrium 
assuming the stiff limit $\sigma \to \infty$ and holding all the other conservative variables in 
$\overline{\mathcal{W}}^0$ constant. 
In our experiments the Newton method applied to eqn. \eqref{eqn.pde.nc.dgO1} with this initial guess typically 
converges to machine zero $(10^{-14})$ after two or three iterations and is robust even for very large values 
of $\sigma$, such as $\sigma=10^{12}$. The resulting cell average $\overline{\mathcal{W}}$ is then used as initial 
guess for the high order space-time solution of eqn. \eqref{eqn.pde.nc.dg4}, i.e. we set $\mathcal{W}^0_h = \overline{\mathcal{W}}$.

\subsection{The Fully Discrete $\PNM$ Schemes} 
\label{sec.ADERNC}

The fully discrete one-step form of the proposed $\PNM$ schemes is derived as follows: we first apply 
the operator $\left< \Phi_k, {\cdot} \right>_{Q_i}$ to PDE  \eqref{eqn.pde.nc} and obtain  
\begin{equation}
\label{eqn.pde.nc.gw1} 
 \left< \Phi_k, \frac{\partial}{\partial t} W \right>_{Q_i} 
    + \left< \Phi_k, \nabla \cdot \tens{F}(W) \right>_{Q_i} = \left< \Phi_k, S(W) \right>_{Q_i}. 
\end{equation}
For the first term in Eqn. \eqref{eqn.pde.nc.gw1} we approximate $W$ with $u_h$ from the space $V_h$ and perform integration by parts in time. Note that the $\Phi_k$ do not depend on time and therefore their time derivatives vanish. For all the other terms 
in  Eqn. \eqref{eqn.pde.nc.gw1} the vector $W$ is
approximated by the solution $\mathcal{W}_h$ of the local
space-time discontinuous Galerkin predictor of section
\ref{sec.galerkin}. 
Since $\mathcal{W}_h$ will usually exhibit jumps at the element boundaries, 
we introduce a numerical flux to resolve these jumps. 
We hence obtain the following family of fully discrete one-step $\PNM$ 
schemes for PDE \eqref{eqn.pde.nc}: 
\vspace{-7mm}
\begin{eqnarray}
\label{eqn.ader.nc.final} 
 \left[ \Phi_k, u_h^{n+1} \right]_{Q_i}^{t^{n+1}} -
 \left[ \Phi_k, u_h^{n  } \right]_{Q_i}^{t^{n}}  
    - \left< \tens{F}_h, \nabla \Phi_k \right>_{Q_i \backslash \partial Q_i}   \nonumber \\
    + \left\{ \Phi_k,  \mathcal{\tens{G}}_{i+\halb}(\WL, \WR) \cdot \vec n 	\right\}_{\partial Q_i}
    = \left< \Phi_k, S(\mathcal{W}_h) \right>_{Q_i}, 
\end{eqnarray}
where $\WL$ denotes the boundary extrapolated data from
within element $Q_i$ and $\WR$ denotes the boundary
extrapolated data from the neighbor, respectively. 
%
%
In the test sections of this paper, we use the Rusanov flux for $\mathcal{G}_{i+\halb}$, which in the case of 
the resistive relativistic MHD equations becomes particularly simple. Due to the presence of the full Maxwell 
equations, whose maximum eigenvalue is the speed of light, i.e. $\lambda_{\max}=1$, it reduces to 
\begin{equation}
 \label{eqn.viscous.rusanov.flux}
  \mathcal{\tens{G}}_{i+\halb}(\WL, \WR) \cdot \vec n = \halb \left( \tens{F}(\WR) + \tens{F}(\WL) \right) \cdot \vec n  - 
                                 \halb \left( \WR - \WL \right). 
\end{equation}
As an alternative, we also propose the following strategy, which gives slightly better results for the hydrodynamic
quantities: For the evolution of the hydrodynamics, Eq.~(\ref{fluid1})-(\ref{fluid5}), one can use the HLL flux  
\be
 \label{eqn.viscous.hll.flux}
  \mathcal{\tens{G}}_{i+\halb}(\WL, \WR) \cdot \vec n = 
\frac{ \left( a^+\tens{F}(\WR) + a^-\tens{F}(\WL) \right)
  \cdot \vec n  - a^+ a^-
  \left( \WR - \WL \right)}{a^+ + a^-}
\ .
\ee
with 
\be
\label{eq:apm}
a^+ = \mathrm{max}\{0, \lambda^-_f, \lambda^+_f\}, \qquad  a^- = \mathrm{max}\{0,-\lambda^-_s,-\lambda^+_s\}, 
\ee

where $\lambda_f$ and $\lambda_s$ denote the fastest and the slowest of the ideal MHD magnetosonic speeds 
along the direction of the flux, and computed through the exact or approximate solution of the corresponding 
quartic as in \cite{delZanna2007}.

For a quadrature-free implementation that requires only the solution of one Riemann problem per space-time 
element interface we refer the reader to \cite{Dumbser2008}.

\section{Numerical Test Cases}
\label{sec.appl}

In this section we present some of the test cases of Palenzuela et al. \cite{Palenzuela2009} who used a second order 
accurate TVD scheme with IMEX Runge-Kutta
time-integration on Cartesian meshes. 
In the rest of the
section we use schemes of order better than two in space and time on unstructured triangular meshes and the 
constant $\kappa$ in the equations (\ref{divE}) and (\ref{divB})
for the divergence cleaning is set equal to unity in all tests.  

\subsection{Large Amplitude Alfv\'en Wave}
\label{sec.convergence2d} 

\begin{figure}[!ht]
\begin{center}
\begin{tabular}{ll}
\includegraphics[width=0.48\textwidth]{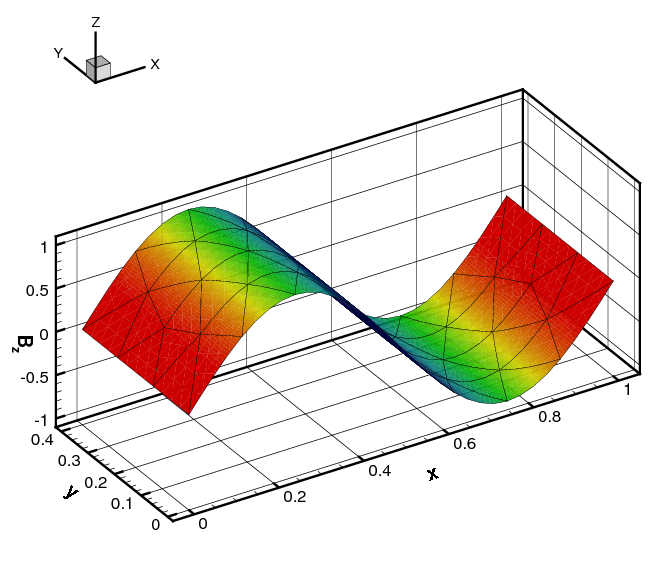} & 
\includegraphics[width=0.48\textwidth]{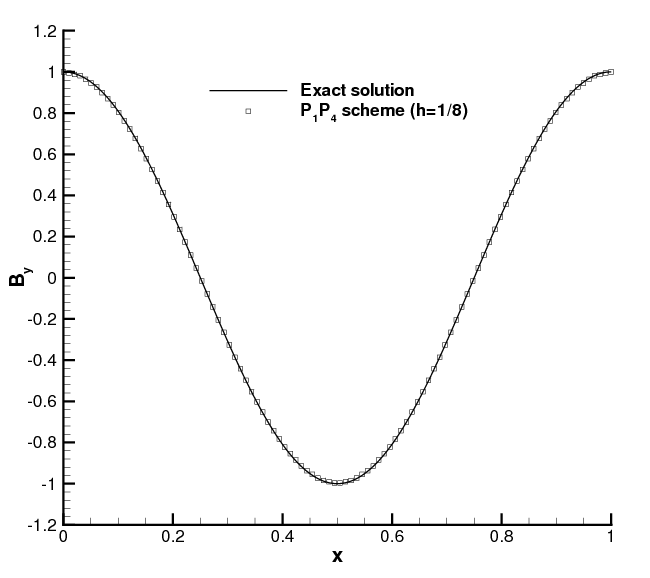} 
\end{tabular}
\caption{Large Amplitude Alfv\'en Wave. 
         Left panel: very coarse unstructured triangular mesh ($h=1/8$) used for the fifth order $P_1P_4$ scheme and surface 
         plot of the quantity $B_z$ at the final time $t=2.618033988$. 
         Right panel: Comparison of exact and numerical solutions for $B_y$ at the final time obtained with the $P_1P_4$ 
         scheme on the very coarse mesh. A cut along the line $y=0$ is shown, evaluating the reconstructed polynomials 
         on 100 equidistant points.}
\label{fig.AlfvenWave}
\end{center}
\end{figure}

This smooth unsteady test case with exact analytical solution was introduced for the ideal relativistic MHD equations 
by Del Zanna et al. \cite{delZanna2007} and was solved for the first time on unstructured triangular meshes with high order
$\PNM$ schemes in \cite{Dumbser2008}. Since the resistive MHD equations tend asymptotically to the ideal ones in the stiff limit ($\sigma \to \infty$), this is an ideal test case to assess the accuracy of our scheme in the stiff limit of the governing PDE  system. \\
The test case consists of a periodic Alfv\'en wave whose initial condition at $t=0$ is chosen to be $\rho=p=1$, $\vec B = B_0 \, (1, \cos\left(k x \right), \sin \left(kx \right) )^T$, 
$\vec v = -v_A/B_0 \, \cdot (0,B_y,B_z)^T$, $\vec E = -\vec v \times \vec B$ and $\phi=\psi=q=0$. 
We furthermore use the parameters $k=2 \pi$, $\gamma=\frac{4}{3}$ and $B_0=1$, hence the advection speed of the Alfv\'en wave in 
$x$-direction is $v_A = 0.38196601125$ (see \cite{delZanna2007} for a closed analytical expression for $v_A$). 
The 2D computational domain is $\Omega = [0;1] \times [0;0.4]$ with four periodic boundary conditions, and 
the final time corresponding to an entire advection
period is $t=1/v_A=2.618033988$. 
The initial condition represents the exact reference
solution to be compared with our numerical one.
Since this test case was constructed for the \textit{ideal} relativistic MHD equations, we have to use a rather  stiff value for the conductivity ($\sigma=10^7$) in the resistive case to reproduce the ideal equations asymptotically. For the fifth order $P_1P_4$ scheme this has shown to be not enough to get the full order of accuracy, hence in this case we even use  $\sigma=10^8$. In all our computations a \textit{constant} Courant number of $\CFL=0.5/(2N+1)$ is used. \\
A representative unstructured triangular mesh is visible
in the left panel of Fig. \ref{fig.AlfvenWave} together with a surface plot 
of the quantity $B_z$. In the right panel we compare the exact solution after one period with the numerical one obtained on a very  coarse mesh of 8 triangles on the $x$-axis using the $P_1P_4$ scheme. For this purpose, the reconstructed fourth degree polynomials  are evaluated at the final time on 100 equidistant points along the $x$-axis in order to make use of the high order polynomial sub-cell resolution contained in each element. We emphasize the excellent agreement  with the exact solution even on this very coarse mesh. Note that with the TVD scheme used in \cite{Palenzuela2009} there were  clearly visible errors even on a fine mesh using 50 points along the $x$-axis. \\ 
Table \ref{tab.resrmhd.conv} shows the errors and the
orders of convergence measured in the $L^2$ norm for the flow variable $B_y$. 
The number $N_G$ denotes the number of triangle edges
along the $x$-axis. 
We stress that the $P_1P_4$ scheme on the very coarse
mesh with $N_G=8$ allows to achieve an accuracy higher 
than the $P_0P_2$ scheme on the finest mesh with
$N_G=64$. 
The nominal order of accuracy $M+1$ has been reached for
all $\PNM$ schemes under consideration.  
In \cite{Palenzuela2009} it was reported that when using
IMEX Runge-Kutta schemes for time-discretization the
authors encountered  problems with the convergence rates
for the relaxed variables, i.e. for the electric field
that suffers from the presence of the stiff source terms.
With our local
space-time Galerkin predictor method, where nonlinear
flux and source term are fully coupled in the predictor
stage and where the \textit{optimal} local space-time
polynomial distribution is found due to the Galerkin
orthogonality property, such problems have not been 
encountered.
Therefore and for the sake of completeness, we show the
convergence rates for the relaxed variable $E_y$ in Table
\ref{tab.resrmhd.stiff} for the schemes $P_0P_2$,
$P_0P_3$ and $P_1P_4$. We deduce from the results of Table 
\ref{tab.resrmhd.stiff} that the nominal order of accuracy 
is reached even for the electric field, which contains the 
stiff source term. This confirms the results already presented 
in \cite{DumbserEnauxToro}, where uniform order of accuracy 
in space and time was found in the stiff as well as in the 
non-stiff case.  

\begin{table}
\caption{Large Amplitude Alfv\'en Wave. Convergence study of $\PNM$ schemes from third to fifth order of accuracy. $\sigma=10^7$,  apart from the $P_1P_4$ scheme where $\sigma=10^8$. Errors are computed for variable $B_y$.}
\vspace{0.5cm}
\begin{center}
\renewcommand{\arraystretch}{1.0}
\begin{tabular}{ccc|ccc|ccc}
\hline
\multicolumn{3}{c|}{$P_0P_2$} & \multicolumn{3}{c|}{$P_1P_2$} & \multicolumn{3}{c}{$P_2P_2$} \\
\hline
 $N_G$ &  $L^{2}$ & $\mathcal{O}_{L^2}$ &  $N_G$ &  $L^{2}$ & $\mathcal{O}_{L^2}$ & $N_G$ &  $L^{2}$ & $\mathcal{O}_{L^2}$  \\ 
\hline 
  16  & 1.71E-02 &      &    8  & 9.12E-04 &      &    8  & 8.97E-04 &      \\
  24  & 5.32E-03 & 2.9  &   12  & 2.26E-04 & 3.4  &   12  & 2.92E-04 & 2.8  \\
  32  & 2.26E-03 & 3.0  &   16  & 9.34E-05 & 3.1  &   16  & 1.67E-04 & 1.9  \\
  64  & 2.79E-04 & 3.0  &   24  & 2.53E-05 & 3.2  &   24  & 4.98E-05 & 3.0  \\
\hline  
\multicolumn{3}{c|}{$P_0P_3$} & \multicolumn{3}{c|}{$P_1P_3$} & \multicolumn{3}{c}{$P_1P_4$} \\
\hline
 $N_G$ &  $L^{2}$ & $\mathcal{O}_{L^2}$ &  $N_G$ &  $L^{2}$ & $\mathcal{O}_{L^2}$ & $N_G$ &  $L^{2}$ & $\mathcal{O}_{L^2}$  \\ 
\hline 
  12  & 1.81E-03 &      &    4  & 7.18E-03 &      &    4  & 3.32E-03 &      \\
  16  & 4.52E-04 & 4.8  &    8  & 3.75E-04 & 4.3  &    8  & 2.95E-05 & 6.8  \\
  24  & 7.35E-05 & 4.5  &   12  & 7.91E-05 & 3.8  &   12  & 4.46E-06 & 4.7  \\
  32  & 1.98E-05 & 4.6  &   16  & 2.82E-05 & 3.6  &   16  & 1.07E-06 & 5.0  \\
\hline  
\end{tabular}
\end{center}
\label{tab.resrmhd.conv}
\end{table}
\vspace{1cm}
\begin{table}
\caption{Large Amplitude Alfv\'en Wave. Verification of
the order of accuracy for a variable affected by the
stiff source term. 
We use the quantity $E_y$ and some selected $\PNM$ schemes.}
\vspace{0.5cm}
\begin{center}
\renewcommand{\arraystretch}{1.0}
\begin{tabular}{ccc|ccc|ccc}
\hline
\multicolumn{3}{c|}{$P_0P_2$} & \multicolumn{3}{c|}{$P_0P_3$} & \multicolumn{3}{c}{$P_1P_4$} \\
\hline
 $N_G$ &  $L^{2}$ & $\mathcal{O}_{L^2}$ &  $N_G$ &  $L^{2}$ & $\mathcal{O}_{L^2}$ & $N_G$ &  $L^{2}$ & $\mathcal{O}_{L^2}$  \\ 
\hline 
  16  & 7.66E-03 &      &   12  & 6.09E-04 &      &    4  & 9.43E-04 &      \\
  24  & 1.90E-03 & 3.4  &   16  & 2.11E-04 & 3.7  &    8  & 1.22E-05 & 6.3  \\
  32  & 7.75E-04 & 3.1  &   24  & 4.02E-05 & 4.1  &   12  & 2.06E-06 & 4.4  \\
  64  & 9.56E-05 & 3.0  &   32  & 1.14E-05 & 4.4  &   16  & 5.18E-07 & 4.8  \\
\hline  
\end{tabular}
\end{center}
\label{tab.resrmhd.stiff}
\end{table}
\subsection{Self-similar Current Sheet} 
\label{sec.komissarov}
\begin{figure}[!htbp]
\begin{center}
\begin{tabular}{ll}
\includegraphics[width=0.48\textwidth]{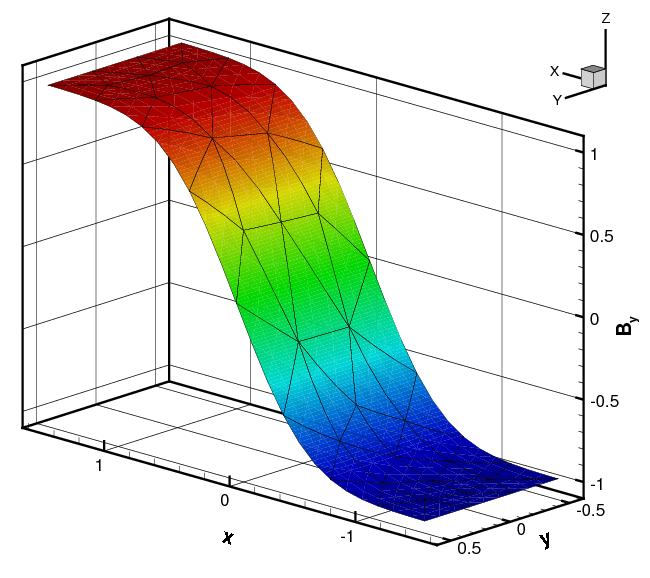} & 
\includegraphics[width=0.48\textwidth]{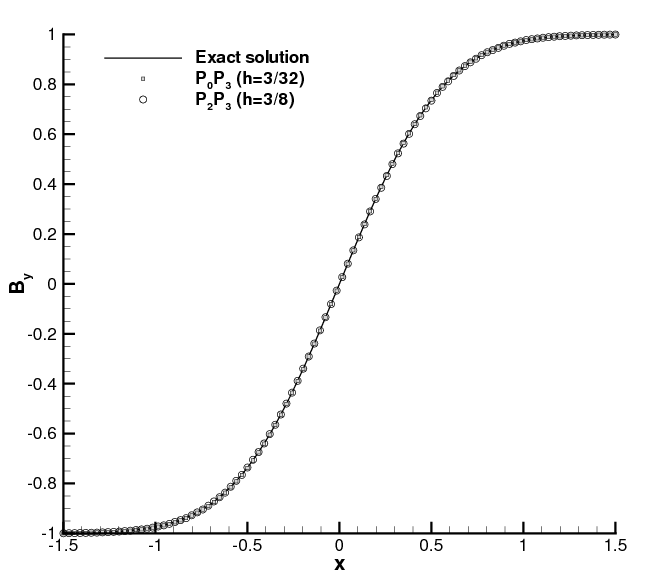} 
\end{tabular}
\caption{Self-similar current sheet. 
         Left panel: unstructured triangular mesh used for the $P_2P_3$ scheme and surface plot of the quantity $B_y$. 
         Right panel: Comparison of exact and numerical solutions at time $t=10$ 
         obtained with two different fourth order $\PNM$ schemes on different meshes. A cut along the line $y=0$ is shown,                   evaluating the reconstructed polynomials on 100 equidistant points.}
\label{fig.komissarov}
\end{center}
\end{figure}

This smooth test case was first proposed by Komissarov et
al. \cite{Komissarov}, it has been presented also in 
Palenzuela et al. \cite{Palenzuela2009} and it provides a
truely resistive test, far from the ideal MHD limit.
It has the following exact analytical solution for the $y$-component 
of the magnetic field:
\begin{equation}
\label{eqn.komissarov.exact}
  B_y(x,t) = B_0 \ \textnormal{erf} \left( \halb
  \sqrt{\frac{\sigma}{t}} x \right) \ ,
\end{equation} 
where ${\rm erf}$ is the error function.
The initial time for this test case is $t=1$ and the initial condition at $t=1$ is given by 
$\rho=1$, $p=50$, $\vec E = \vec v = 0$ and $\vec B =
(0,B_y(x,1),0)^T$. 
We choose 
$\gamma = \frac{4}{3}$ and $B_0 = 1$. The conductivity is chosen as $\sigma = 100$, which means a
moderate resistivity. The problem is solved with two different fourth order $\PNM$ schemes on the 
two-dimensional computational domain $\Omega = [-1.5;1.5] \times [-0.5;0.5]$, where we impose periodic
boundary conditions in $y$-direction and Dirichlet boundary conditions consistent with the initial 
condition in $x$-direction. The first scheme is a pure finite volume method ($P_0P_3$) using the 
component-wise WENO reconstruction proposed in \cite{DumbserKaeser06b}, running on a mesh with $h=3/32$, 
which corresponds to an equivalent one-dimensional resolution of 32 points. The second scheme is the 
$P_2P_3$ method which is part of the new intermediate class of numerical schemes discovered in 
\cite{Dumbser2008}, running on a very coarse mesh with $h=3/8$, i.e. using only 8 points in the one-dimensional
case. The mesh is depicted on the left panel of Fig. \ref{fig.komissarov} together with a surface plot of the 
magnetic field in $y$-direction. Both numerical solutions are compared at time $t=10$ with the exact
solution given by Komissarov et al. \cite{Komissarov} and Palenzuela et al. \cite{Palenzuela2009} on the
right panel of Fig. \ref{fig.komissarov}. We note an excellent agreement with the exact solution and underline that 
the use of high order methods in \textit{space and time} allows us to use very coarse meshes, compared to standard second order TVD schemes.    
%

\subsection{Shock Tube Problems} 
\label{sec.shocktubes}

\begin{figure}[!htbp]
\begin{center}
\begin{tabular}{lcr}
\includegraphics[width=0.31\textwidth]{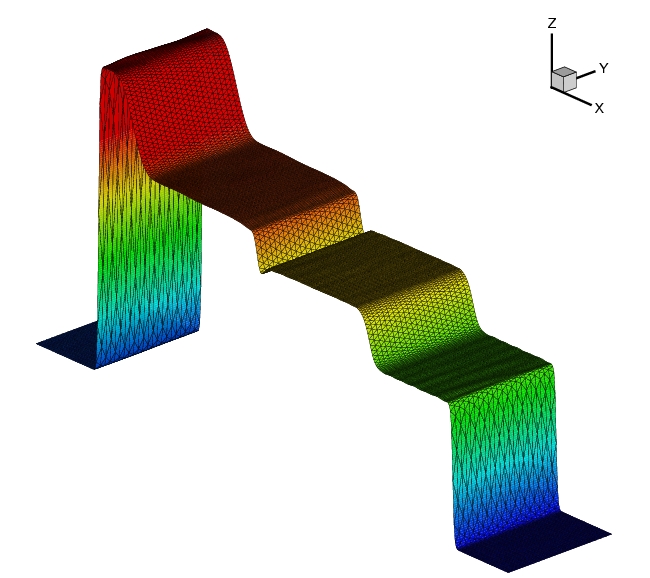} & 
\includegraphics[width=0.31\textwidth]{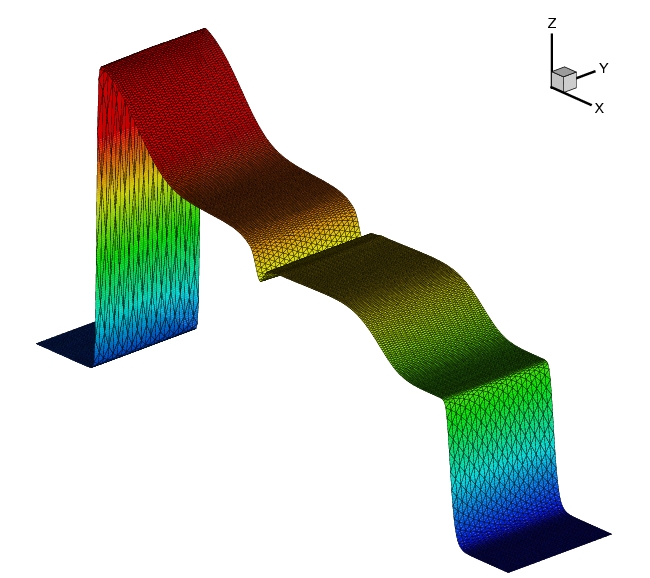} & 
\includegraphics[width=0.31\textwidth]{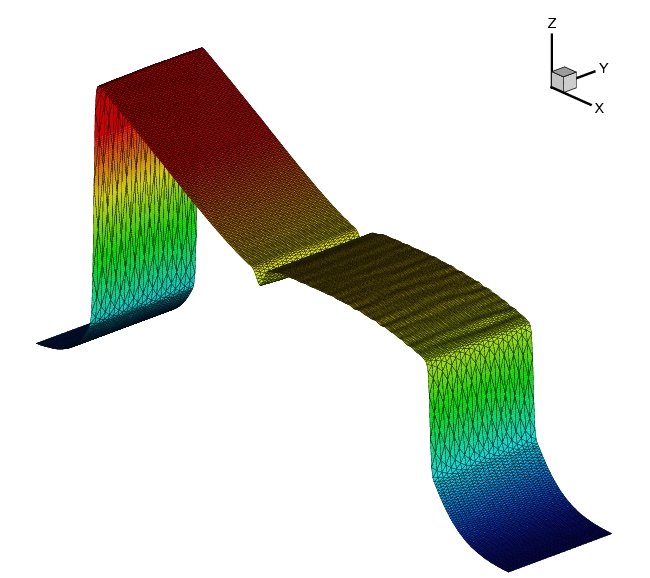}   
\end{tabular}
\caption{RRMHD shock tube test problem using a $P_0P_2$ WENO scheme and different values for the conductivity $\sigma$. 
         The unstructured triangular mesh is shown together with a surface plot of the density $\rho$. 
         Left panel: $\sigma=10^3$. Middle panel:
         $\sigma=10^2$. Right panel: $\sigma=10$. 
				 A cut along the line $y=0$ is shown at 400 equidistant points.}
\label{fig.RMHD5Mesh}
\end{center}
\end{figure}

\begin{figure}[!htbp]
\begin{center}
\includegraphics[width=0.8\textwidth]{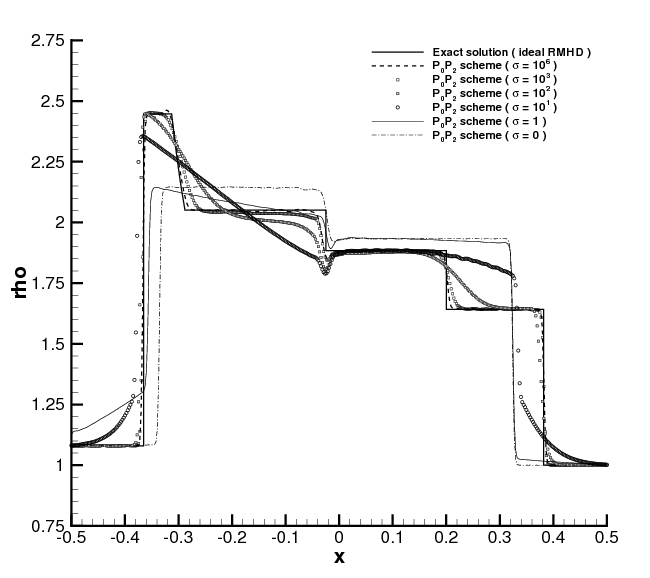} \\ 
\includegraphics[width=0.8\textwidth]{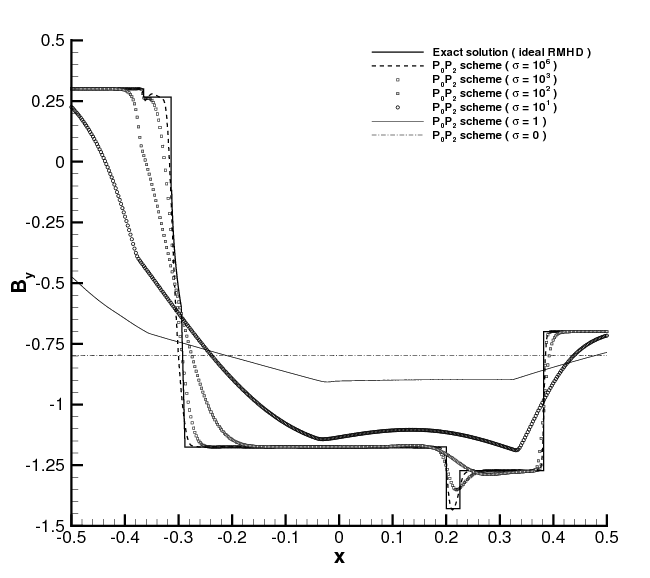}  
\caption{RRMHD shock tube test problem using a $P_0P_2$ WENO scheme and different values for the conductivity $\sigma$. 
         The exact solution is shown for the ideal RMHD
         equations. The density $\rho$ is 
         plotted on the top of the figure and the magnetic field component 
         $B_y$ on the bottom.}
\label{fig.RMHD5}
\end{center}
\end{figure}

In this section we solve the fifth of a series of test problems proposed by Balsara in \cite{BalsaraRMHD}. We solve the RRMHD equations with different values for the conductivity $\sigma$. The initial condition is given by two piecewise constant states separated by a discontinuity at $x=0$. The left and right values for the primitive variables are reported in Table \ref{tab.rmhd.ic}. Furthermore, we set $\vec E = - \vec v \times \vec B$, $\phi=\psi=q=0$ and $\gamma=\frac{5}{3}$. The conductivities in our test cases are chosen as $\sigma=0$, $\sigma=1$, $\sigma=10$, $\sigma=10^2$, $\sigma=10^3$ and $\sigma=10^6$. The computational domain is $\Omega = [-0.5;0.5] \times [0;0.05]$ with periodic boundaries in $y$-direction and Dirichlet boundaries consistent with the initial condition in $x$-direction. We use an unstructured triangular mesh of characteristic size $h=1/400$, which is depicted together with a surface plot of the density $\rho$ in Fig. \ref{fig.RMHD5Mesh}. A cut through the solution along the $x$-axis is shown in Fig. \ref{fig.RMHD5} for all different values of $\sigma$ used in this series of test cases. The exact solution is the one for the ideal RMHD equations, as published in \cite{GiacomazzoRezzolla}. The essential wave structures of the ideal RMHD Riemann problem can be noted for $\sigma=10^3$ or greater. For values below, the resistivity leads to a considerable diffusion of the discontinuities. 

\begin{table}
\caption{Initial states left (L) and right (R) for the
  relativistic MHD shock tube problem. The last column
  reports the final time $t_e$ considered in the
  numerical test.}
\vspace{0.5cm}
\renewcommand{\arraystretch}{1.0}
\begin{center}
\begin{tabular}{cccccccccc}
\hline
 Case  & $\rho$ & $p$ & $u$  & $v$ & $w$ & $B_y$ & $B_z$ & $B_x$ & $t_e$ \\
\hline 
L & 1.08 & 0.95 & 0.4   & 0.3  & 0.2 & 0.3  & 0.3 & 2.0 & 0.55 \\
R & 1.0  & 1.0  & -0.45 & -0.2 & 0.2 & -0.7 & 0.5 & 2.0 &      \\
\hline
\end{tabular}
\end{center}
\label{tab.rmhd.ic}
\end{table}

\subsection{Rotor Problem} 
\label{sec.rotor}

\begin{figure}[!htbp]
\begin{center}
\begin{tabular}{lcr}
\includegraphics[width=0.47\textwidth]{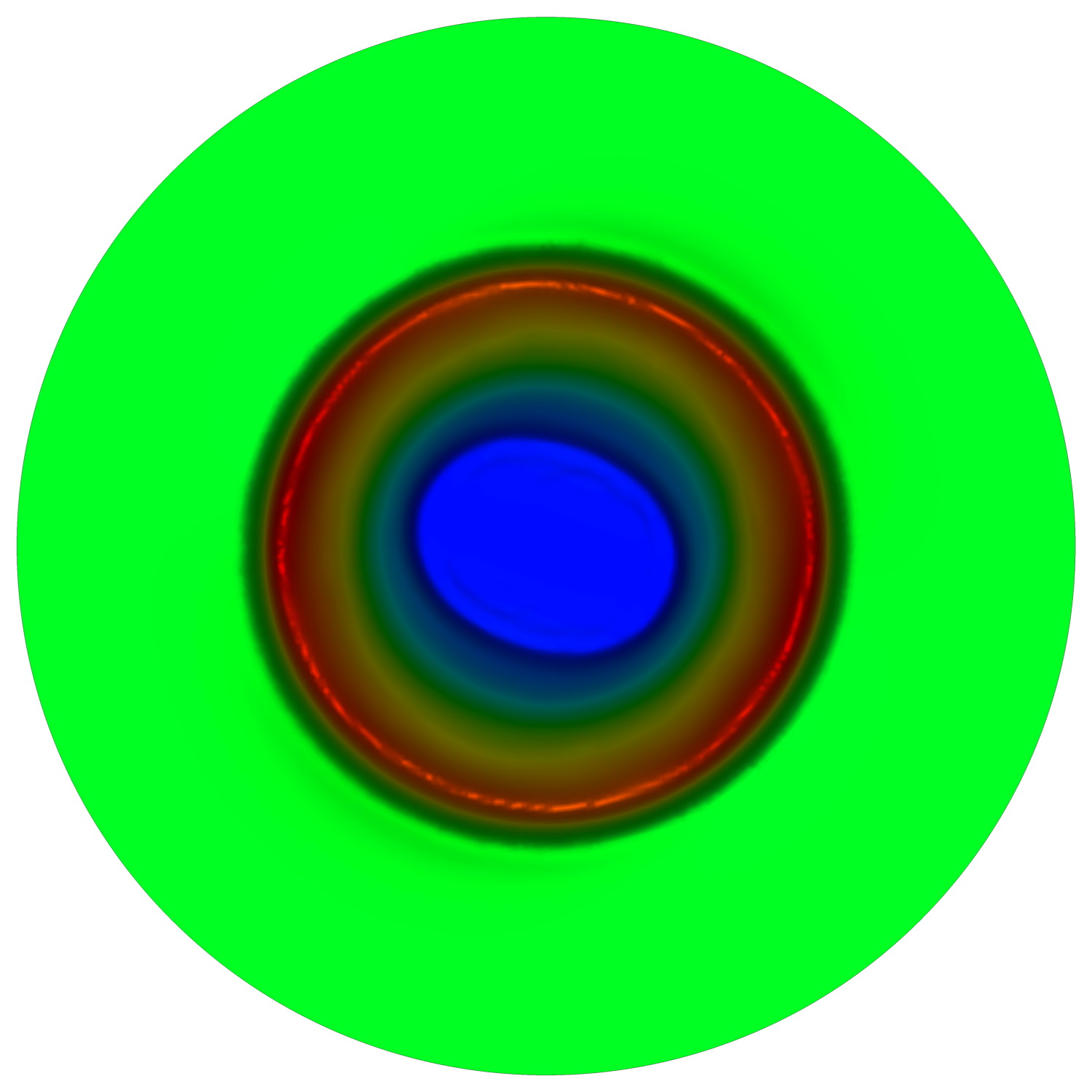}  &
\includegraphics[width=0.47\textwidth]{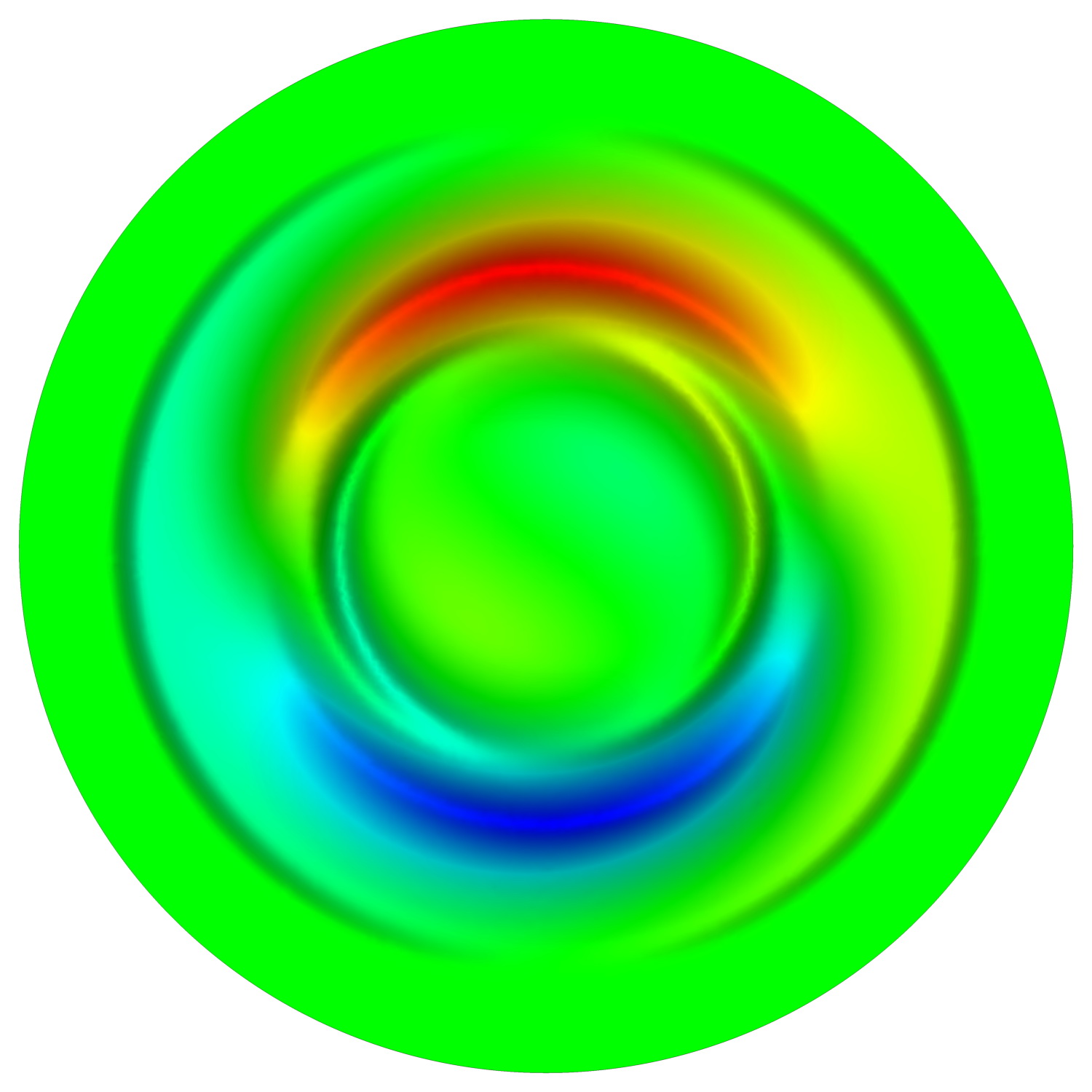} \\  
\includegraphics[width=0.47\textwidth]{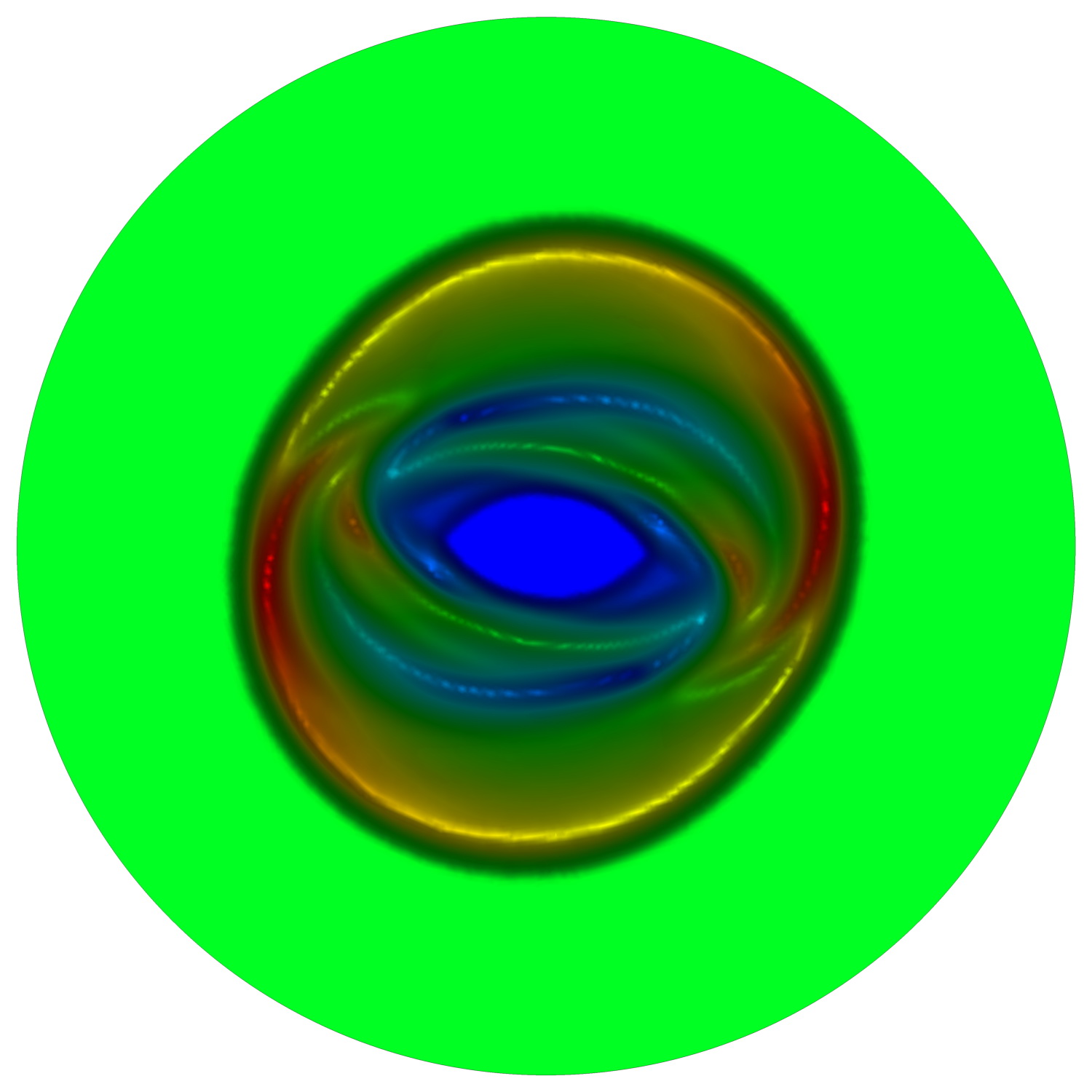}  &
\includegraphics[width=0.47\textwidth]{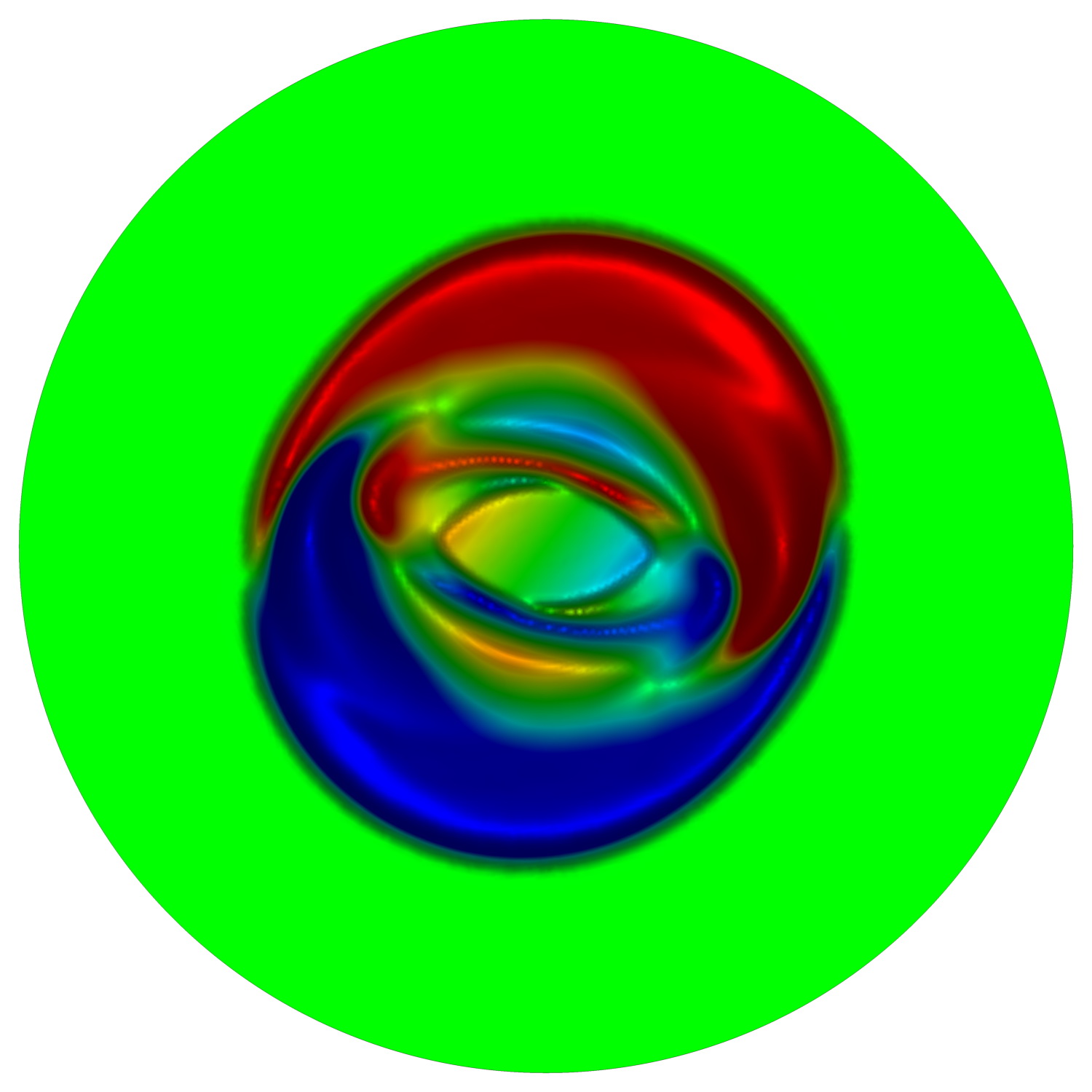} \\  
\includegraphics[width=0.47\textwidth]{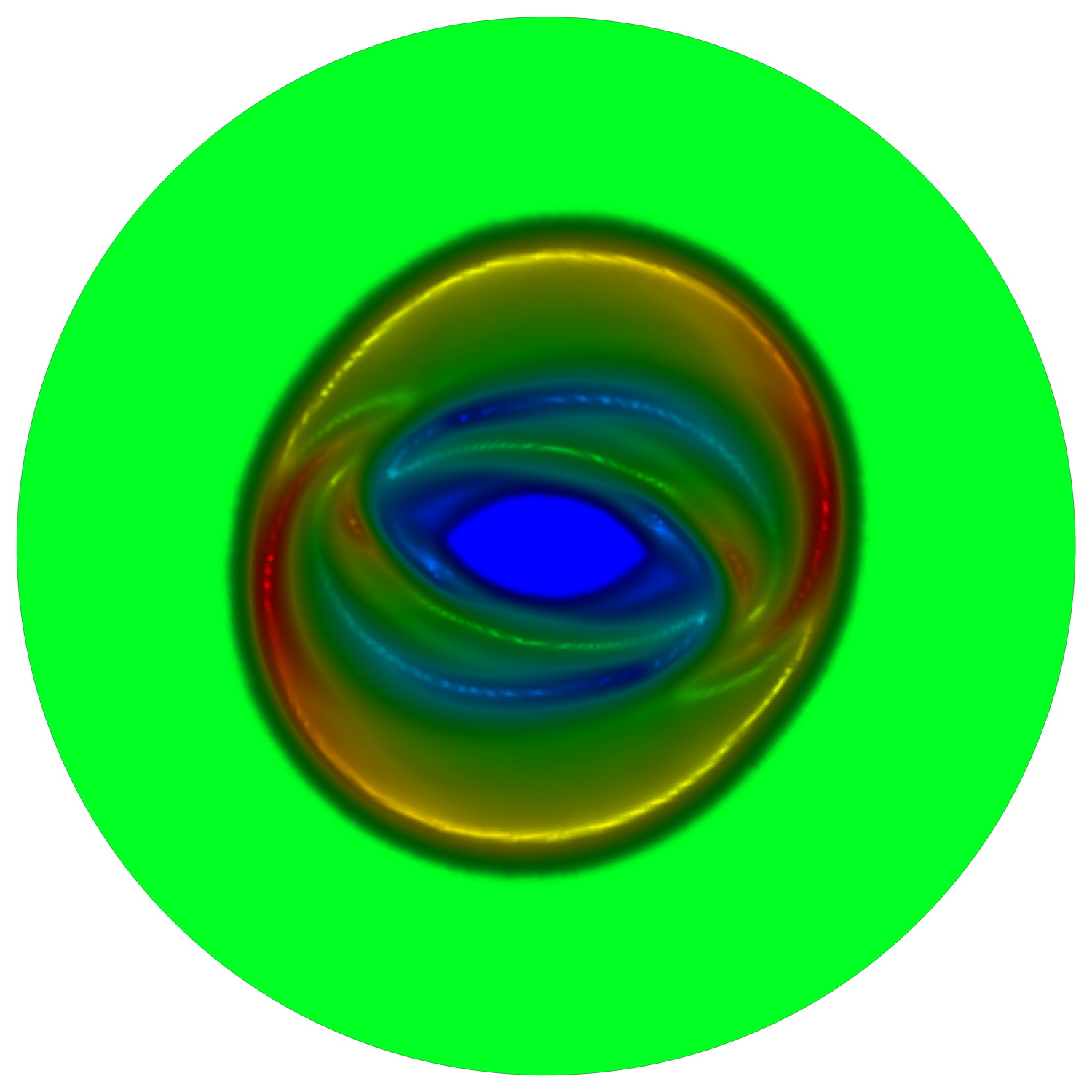}  &
\includegraphics[width=0.47\textwidth]{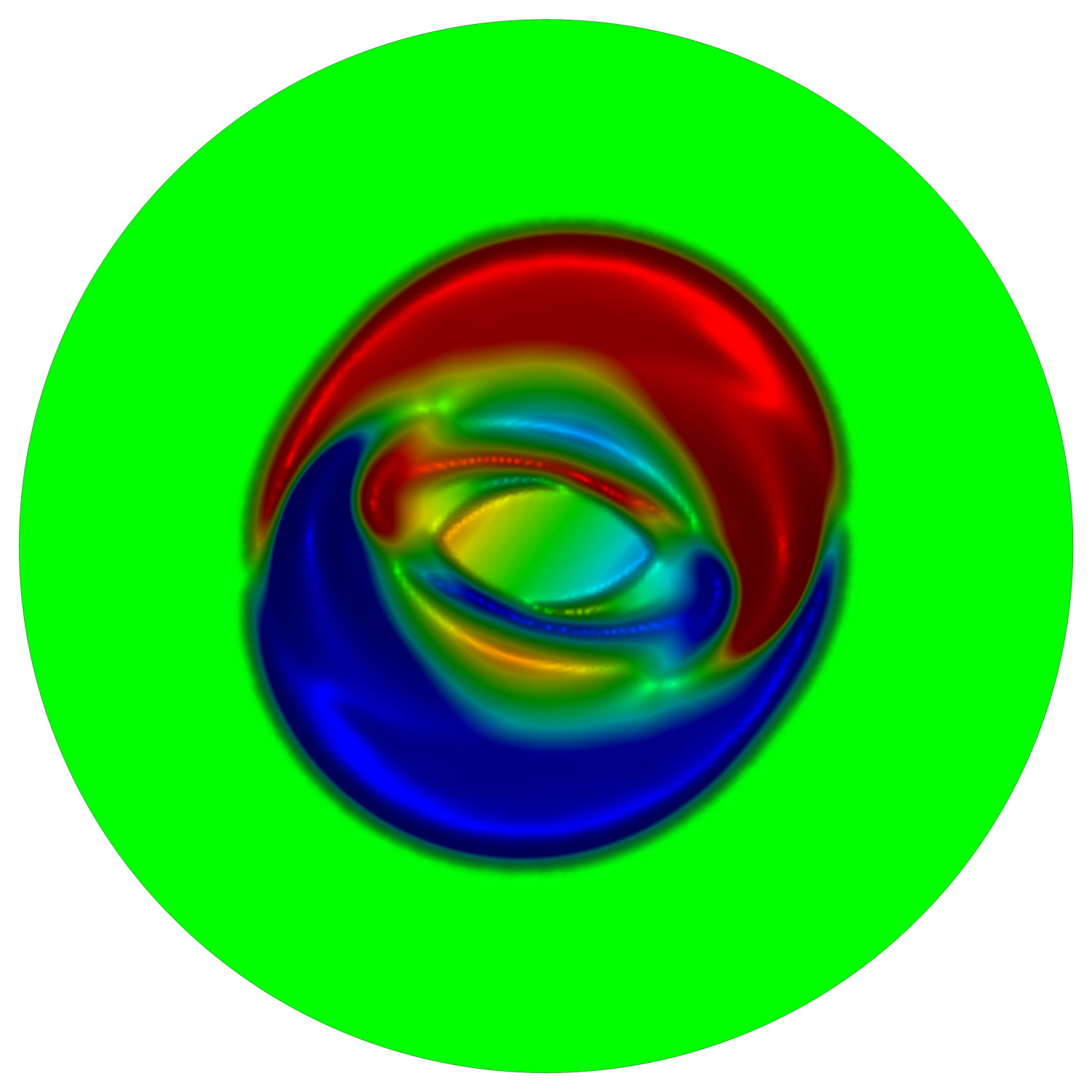}    
\end{tabular}
\caption{Pressure field (left column) and electric field component $E_z$ (right column) for the resistive relativistic 
         rotor problem at time $t=0.3$. 
         Top: $\sigma=10$. Middle: $\sigma=10^5$. Bottom: ideal RMHD.}
\label{fig.rotor}
\end{center}
\end{figure}

In this section we solve a resistive relativistic version of the MHD rotor problem proposed by Balsara and Spicer  \cite{BalsaraSpicer1999}. Our computational setup is a variation of the ideal relativistic MHD rotor test case of Del Zanna  
et al. \cite{delZanna2003}. In contrast to \cite{delZanna2003}, who solved the ideal RMHD equations on a perfectly regular Cartesian mesh, 
we solve this test case in Cartesian coordinates on a \textit{circular} computational domain with radius $R=0.5$ using an \textit{unstructured triangular} mesh with a characteristic mesh spacing of $h=0.004$ towards the center and $h=0.005$ at the 
outer border of the domain, leading to a total number of 72320 triangles. The rotor has a radius of $R_0=0.1$ and is spinning 
with an angular frequency of $\omega_s = 8.5$, leading to a maximal toroidal velocity of $v_\phi=0.85$. The density is $\rho=10$ 
inside the rotor and $\rho=1$ in the outer fluid at rest. The pressure is $p=1$ and the magnetic field is $\vec B = (1, 0, 0)^T$ in the whole domain. The initial electric field is, as usual, computed as $\vec E = -\vec v \times \vec B$. We use a $P_0P_2$ scheme 
with component-wise WENO reconstruction. No taper is
applied to the initial condition, as in
\cite{delZanna2003}, and $\gamma=4/3$ is
used. Transmissive boundary conditions are applied at the
outer boundaries. The computational domain and the
results for the pressure at  time $t=0.3$ are shown in
Fig. \ref{fig.rotor} for different values of the electric
conductivity. We solve the problem with $\sigma=10$ and
$\sigma=10^5$ and, as a reference solution, we also show
the results obtained with the ideal RMHD equations. The
ideal RMHD results agree qualitatively very well with
those obtained with the RRMHD equations using
the larger conductivity $\sigma=10^5$. For the case of a
lower conductivity ($\sigma=10$) one can clearly see that
the wave structure is  completely different, with a
faster moving electric field that is governed directly by
the Maxwell equations and no longer resulting from the
relation $\vec E = -\vec v \times \vec B$ as in the ideal
case. 
\subsection{Orszag-Tang Vortex}
\label{sec.orszagtang} 

\begin{figure}[!htbp]
\begin{center}
\begin{tabular}{lr}
\includegraphics[width=0.47\textwidth]{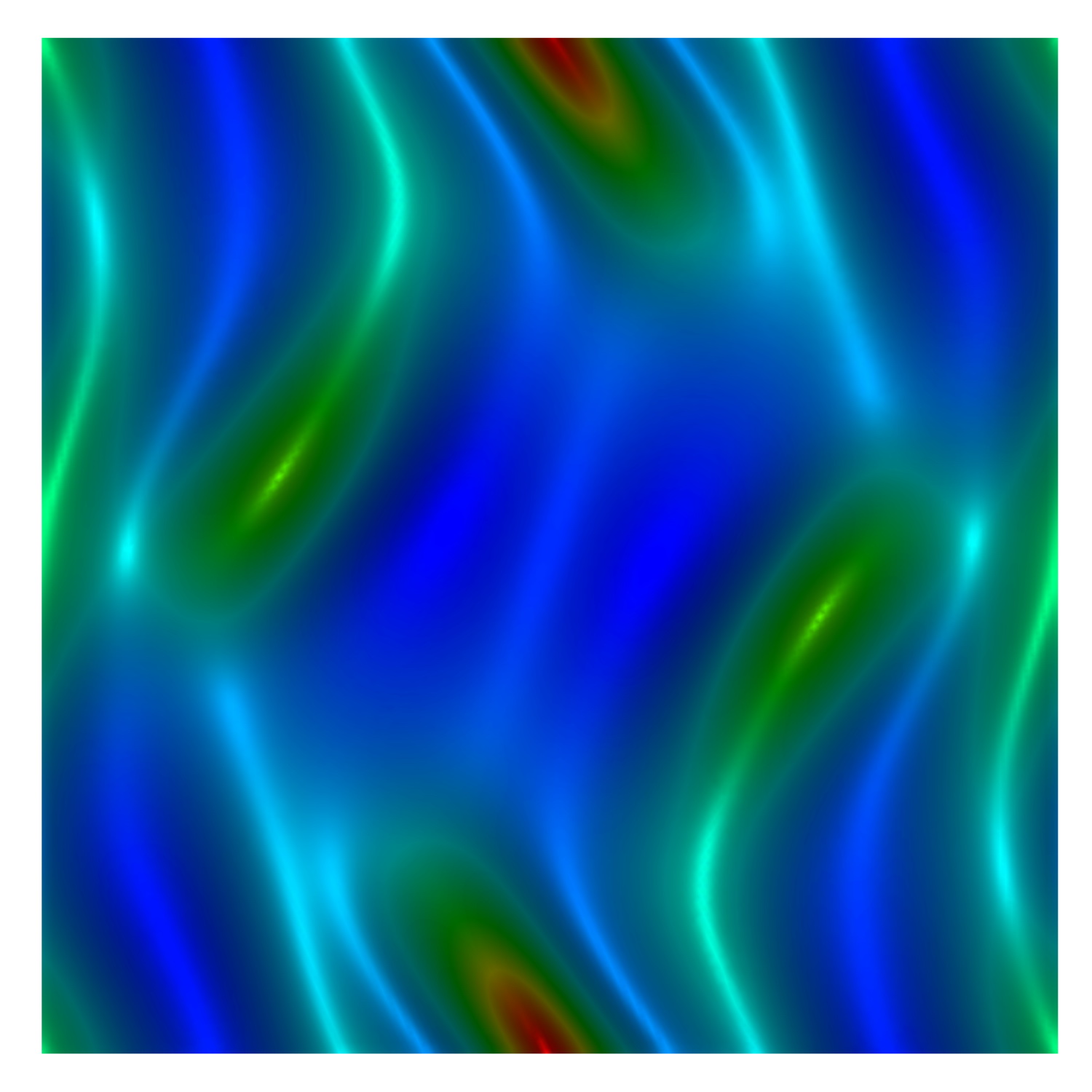} & 
\includegraphics[width=0.47\textwidth]{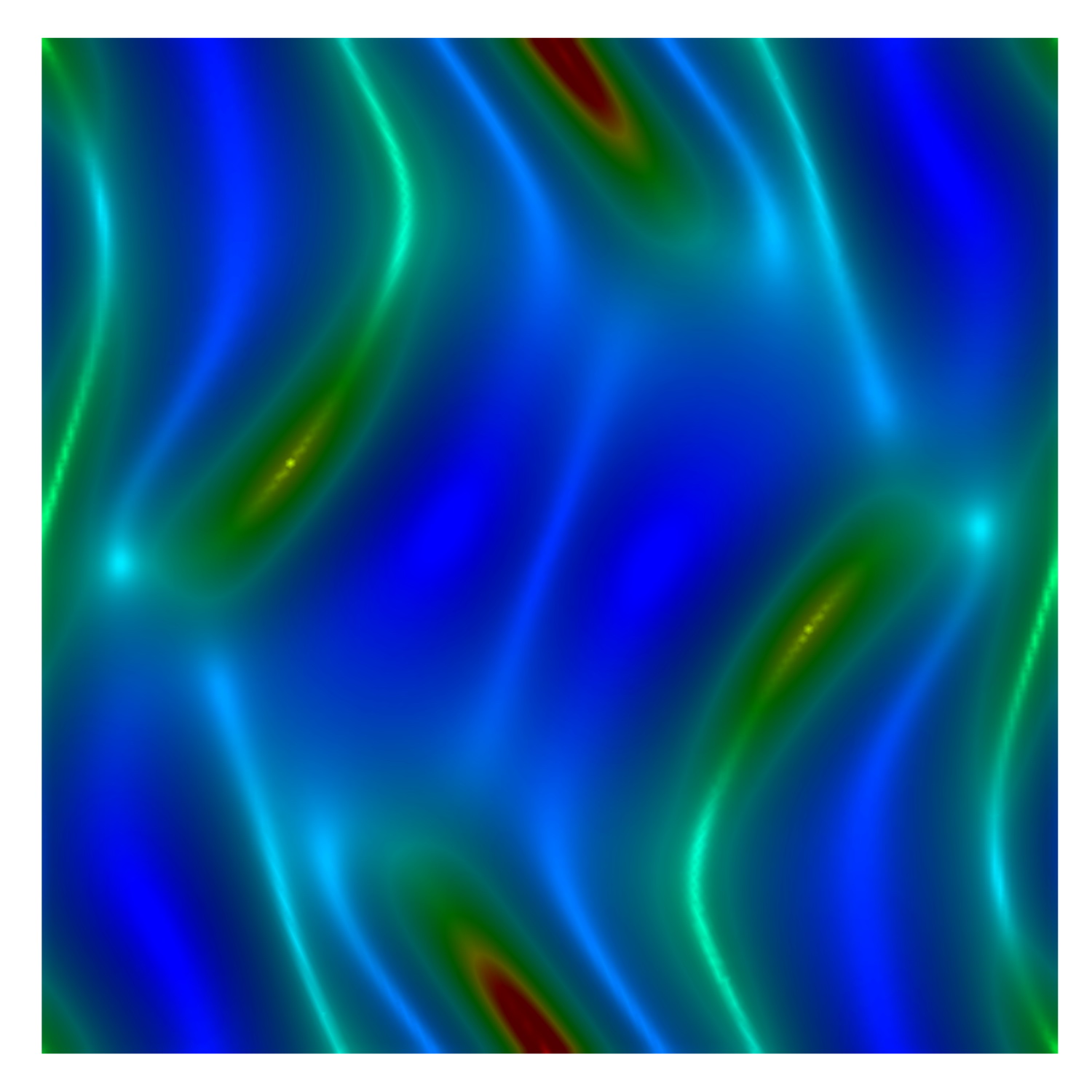} \\ 
\includegraphics[width=0.47\textwidth]{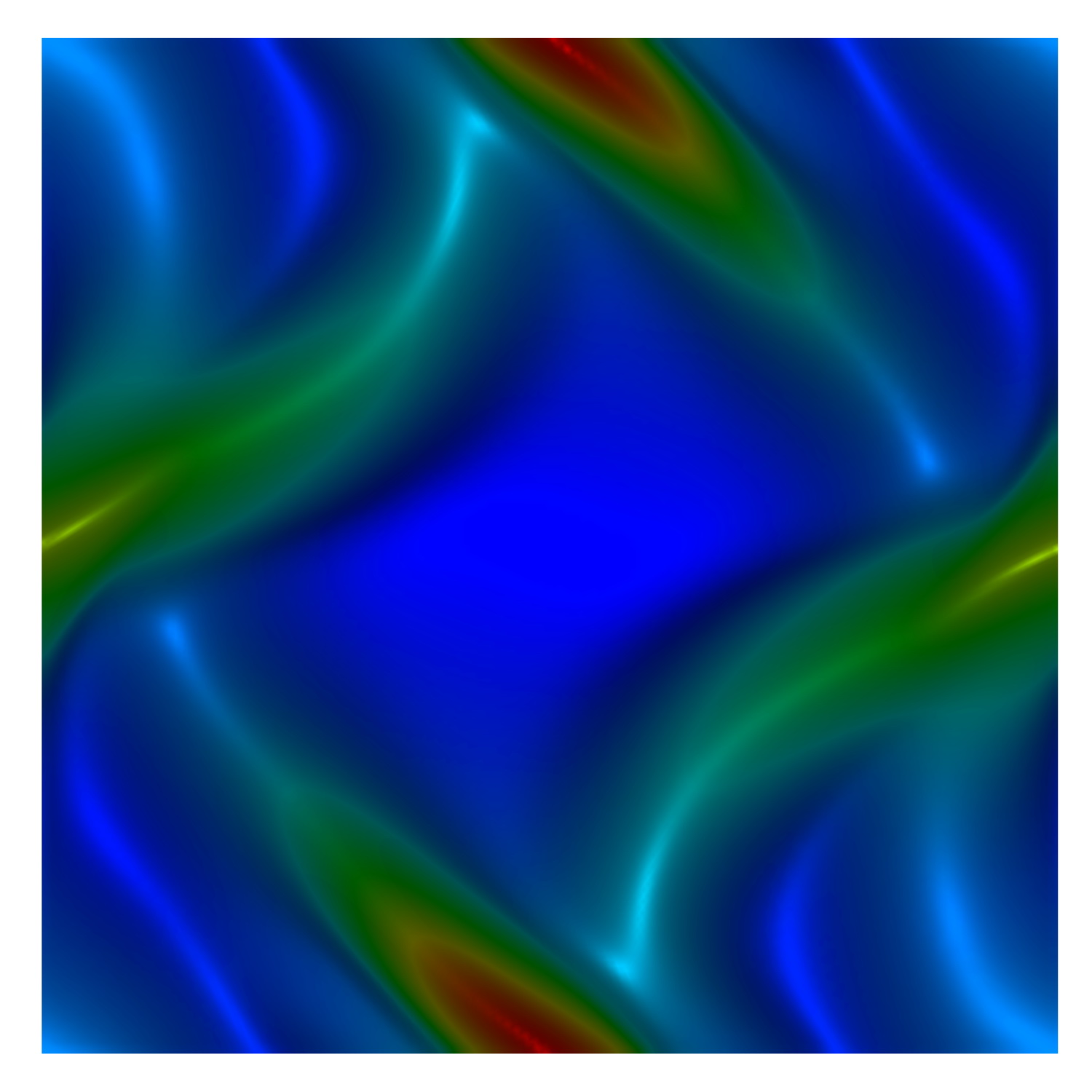} & 
\includegraphics[width=0.47\textwidth]{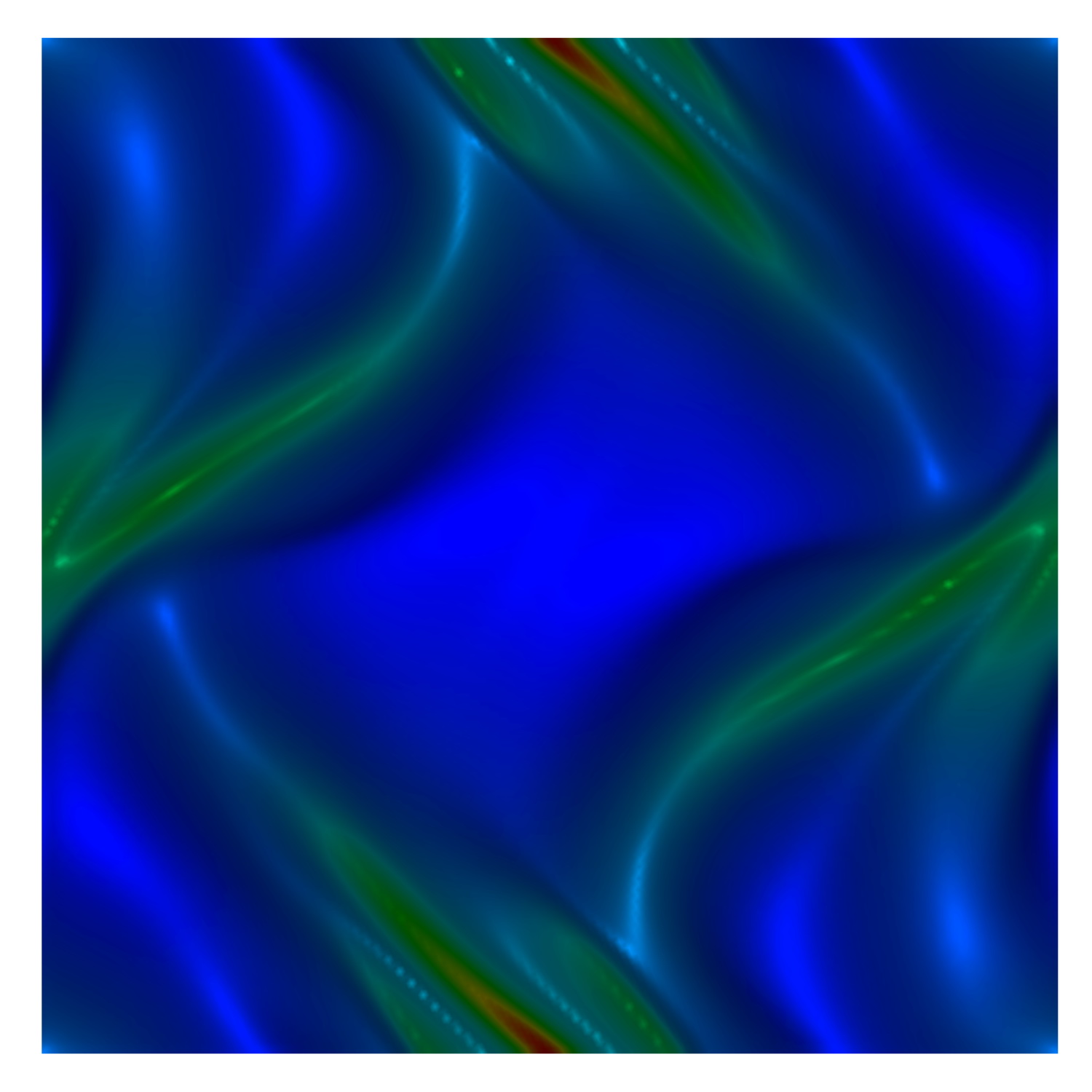} \\ 
\includegraphics[width=0.47\textwidth]{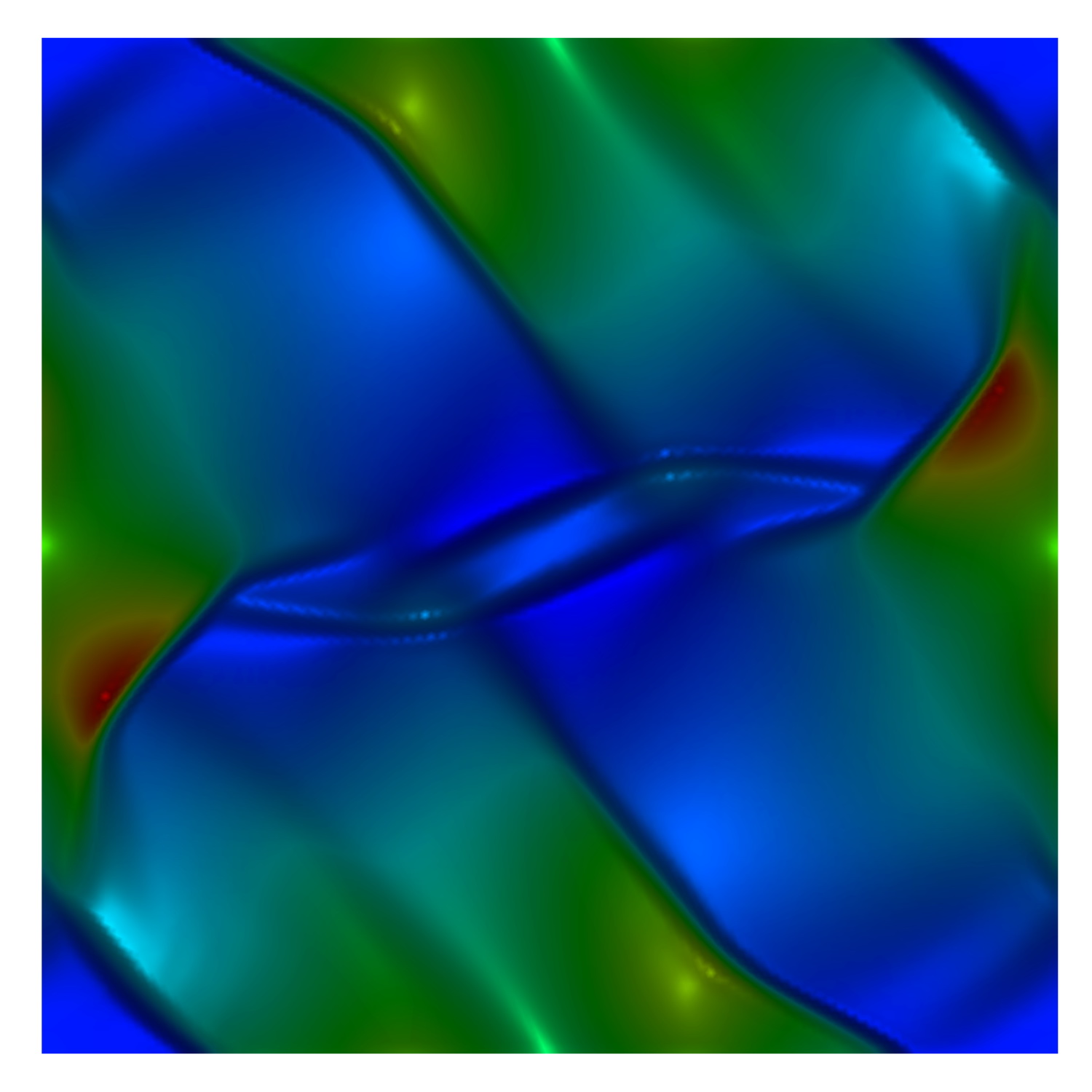} & 
\includegraphics[width=0.47\textwidth]{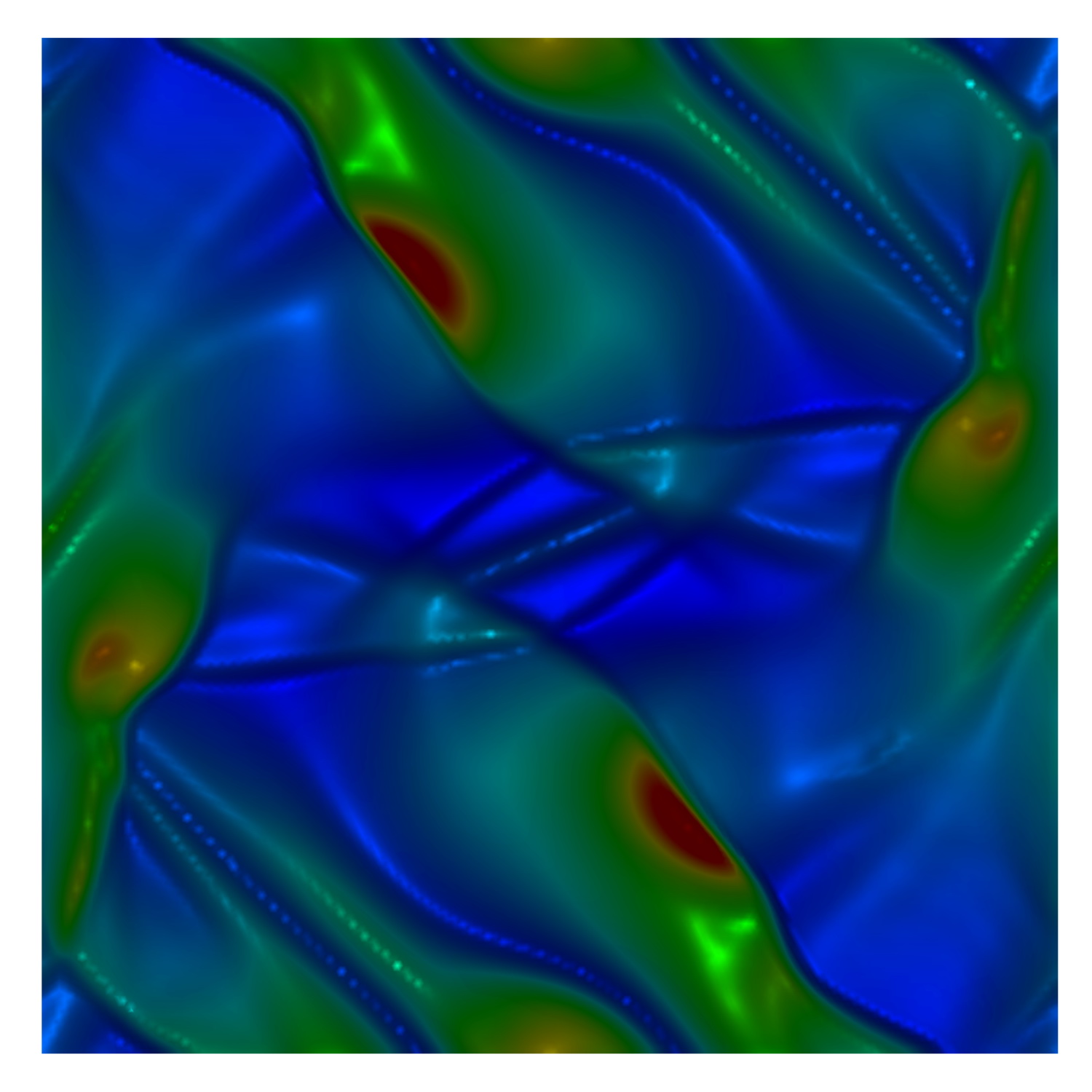}   
\end{tabular}
\caption{Pressure field for the resistive relativistic Orszag-Tang vortex problem at times $t=0.5$, $t=2.0$ and $t=4.5$ (from top to bottom). Left column: $\sigma=10$. Right column: $\sigma=10^3$.}
\label{fig.otang}
\end{center}
\end{figure}

The last of this series of test cases is a resistive
relativistic analogous 
of the Orszag-Tang vortex problem \cite{OrszagTang} studied extensively in \cite{DahlburgPicone,PiconeDahlburg,JiangWu}.   
The computational domain is $\Omega = \left[0;2\pi\right]^2$. The initial condition of the problem is given by  
\begin{equation}
  \left( \rho,u,v,p,B_x,B_y \right) = \left( 1, -\sin(y), \sin(x), 1, - B_0 \sin(y), B_0 \sin(2x) \right), 
\end{equation}
with $w=B_z=0$ and $\gamma = \frac{4}{3}$. The problem is solved up to $t=4.5$ using a $P_0P_2$ scheme with component-wise 
WENO reconstruction on an unstructured triangular mesh with 55292 elements ($h=\frac{1}{25}$). The results are shown for the  pressure in Fig. \ref{fig.otang} for times $t=0.5$, $t=2.0$ and $t=4.5$ using two different values for the conductivity, $\sigma=10$ and $\sigma=10^3$. As in the original problem \cite{OrszagTang}, the smooth sinusoidal initial condition evolves in time to form complex shock dominated structures for the large value of the conductivity. For small conductivities, much less waves are present due to the diffusion caused by the electric resistivity.  

\section{Conclusions}
\label{sec.conclusion}

In this paper we have solved the resistive relativistic
magnetohydrodynamics equations using the class of methods 
introduced in Dumbser et al. \cite{Dumbser2008} and named
$\PNM$ schemes. The equations present source terms that
are potentially stiff when the ideal limit of infinite
conductivity is recovered. As such, they are naturally
accounted for through the application of the local 
space-time discontinuous Galerkin predictor, originally 
deviced in \cite{DumbserEnauxToro}. 

To our knowledge, the computations presented in this
paper are the first better than second order accurate
simulations {\em in space and time} ever done for the stiff
limit of the RRMHD equations and the results obtained point to favour 
higher order methods over standard second order TVD schemes.    
In particular, the accuracy that can be achieved with
high order $\PNM$ schemes on very coarse meshes 
makes them promising tools for simulations of physical processes 
that require high computational resources, such as a large 
class of time dependent problems involving magnetic reconnection 
in astrophysical context. Further directions of future 
improvement are represented by the generalization of the 
scheme into full general relativity
as well as the inclusion of more complex Ohm's laws.

\section{Acknowledgments}
The research presented in this paper was financed by the 
\textit{Deutsche Forschungsgemeinschaft} (DFG) by the grant 
\textit{DFG Forschungsstipendium} (DU 1107/1-1). 
This work was also partially supported by COMPSTAR, an ESF 
Research Networking Programme, and by the DFG SFB/Transregio~7.

The authors would also like to thank Bruno Giacomazzo and 
Luciano Rezzolla for  providing the exact reference solutions 
for the Riemann problem of the relativistic MHD equations. 
OZ also thanks Carlos Palenzuela for helpful discussions.


\bibliography{ResRMHD}
\bibliographystyle{plain}

\end{document}